# Slip topology of steady flows around a critical point: Taking the linear velocity field as an example

Wennan Zou [*1], Jian He[1]

1. Institute for Advanced Study / Institute of Fluid Mechanics, Nanchang University, Nanchang 330031, China

*Authors to whom correspondence should be addressed: zouwn@ncu.edu.cn

**ABSTRACT:** The flow of real fluids is the aggregation of the motion of fluid particles when the fluid is conceived to be made up by an infinite number of particles. As an alternative of this conventional model, fluid motion could be understood as the slip of fluid layers with a molecular scale over each other, where the slip structures of fluid and their associated small-scale motion are characterized by an axial-vector-valued differential 1-form, called the vortex field. In this paper, in the case of steady flows we define the swirling degree of the velocity field at a point, and further the swirl field of the steady flow, to study the slip topology of fluid or the local streamline pattern around the critical point. The linear velocity field in the right real Schur form is used to carry out detailed analyses around the isolated critical point. Theoretical deduction and numerical test unveil the connection between the swirling degree and the swirl field, greatly make clear the topological property of slip structures of fluid in steady flows, especially in three-dimensional space.

**Key Words:** steady flow; swirling degree; streamline pattern; slip topology; critical point

## 1 Introduction

Vortex in reality is a basic phenomenon in fluid flow, which has strong dynamic effect and is also one of the essential characteristics of turbulence. Lugt (1983) presented the definition that a vortex is the rotating motion of a multitude of material particles around a common center, as an equal to the vorticity definition. Foss and Williams (1990) pointed out "how shall we (the fluid mechanics community) most reliably relate the precisely defined quantities such as 'vorticity' and 'vortex lines' to the seemingly intuitive quantities such as 'a vortex' and 'a large-scale motion'". Robinson (1991) proposed that a vortex exists when instantaneous streamlines mapped onto a plane normal to the vortex core exhibit a roughly circular or spiral pattern, when viewed from a reference frame moving with the center of the vortex core. Many efforts have been made to clarify what is a real vortex, but up to now, it has not become clearer.

Superimposing any inertial velocity on a stable vortex flow will change the streamline pattern, until no vortex can be observed through streamlines. In complex flows, where vortices become unstable, transported or even break into random small-scale motion, it seems an impossible task for us to identify them one by one. Following the traditional understanding, vortex is a feature of fluid motion characterized by the velocity field, must be invariant under the Galilean transformation due to the requirement of objectivity (Jeong and Hussain 1995; Chakraborty *et al*. 2005; Tian *et al*. 2018). If we get rid of this thinking constraint, and regard vortex as a structure of fluid in motion, the stable vortex is always there, let alone the observer in what inertial reference frame (Zou *et al*. 2021). From this viewpoint, people no longer feel confused that the studies of flow patterns stemming from the critical point theory (see Dallmann, 1983; Perry and Chong, 1987 and the literature therein) depend on the observer (Chong *et al*., 1990).

Vollmers *et al*. (1983), Dallmann (1983) and Chong *et al*. (1990) put forward the definition that a vortex core is a region of space where velocity gradient has complex eigenvalues. Zhou *et al*. (1999) presented the spiral streamline around the critical point happens when velocity gradient has complex eigenvalues. Zou *et al*. (2021) insist on maintaining the critical point as a basic character of flow structures and pointed out the dual directivity of spiral streamline pattern in virtue of the study on the right eigen-representation of velocity gradient. Following another line of thought, that is, breaking through the Helmholtz decomposition, Kolář (2004, 2007) proposed a triple decomposition by extracting a so-called "effective" pure shearing motion from the spin tensor; Li *et al*. (2014) presented the quadruple decomposition of velocity gradient based on the real Schur form, namely, dilatation/contraction, axial stretch along some axis, plane motion and simple shear; Liu and his co-workers (Liu *et al*. 2016, 2018) strongly questioned the uniqueness of using vorticity to describe vortices, and recommended the Rortex method after putting forward the Ω-method (Liu *et al*. 2019).

Our previous study (Zou *et al*. 2021) established a basic frame of linear velocity field, namely the frame of the right eigen-representation of velocity gradient, and pointed out that the classification based on the parameter set of right eigen-representation is not of topological. These consist of the foundation in the study of this paper. A vortex must have a core with dual directionality: the rotating axis and the extending direction (Fig. 1). The rotating axis reflects the kinematic characteristics of the vortex, while the extension direction reflects the existence characteristics of the vortex. The aim of this paper is to explain the fact that the vortex flow is first of all a curved flow, but the curved flow does not necessarily form a vortex. The latter is the real representative of the vortex core. We first introduce a topological index, called the swirling degree, from the mapping of velocity direction onto the unit sphere, to characterize the streamline pattern around the critical point. Then, by regarding the flowing fluid as a differentiable manifold covered by fluid elements with micro-finite scale and inner orientation, instead of a continuous aggregation of isolated fluid particles without any volume, a geometrical quality, called the swirl field, is constructed with the following properties: (1) it indicates the orientation difference of adjacent fluid elements, (2) it connects the velocities of adjacent fluid elements by covariant differential, (3) it can be expressed by the velocity direction in the steady flows.

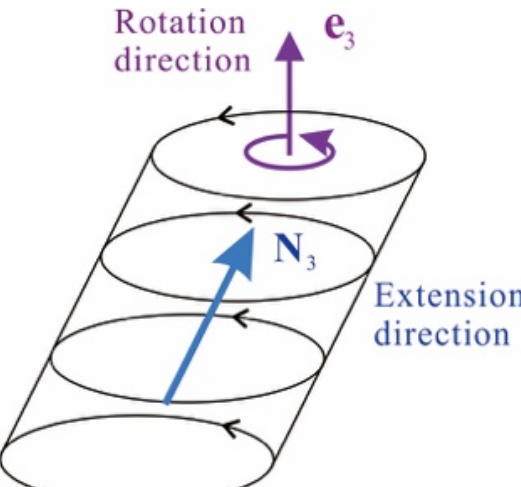

**Figure 1**. Dual directivity of spiral streamline pattern

In this paper, we will investigate the topological property of streamline pattern around a critical point by studying the swirling degree and the swirl field of steady flows, with linear velocity fields as examples. The content is arranged as follows. The concept of swirling degree is introduced in §2, and some formulae of linear velocity field are presented for the use in next sections. In §3, the swirl field is defined in a more generalized background, and its expression in steady flows is derived in terms of the orientation matrix. Beginning with the simple two-dimensional (2D) linear velocity fields in §4, we study the topology of streamline pattern around the critical point, and then the formulae of topological indices in three-dimensional (3D) velocity fields are derived and analyzed in details. In §5, the properties around an isolated critical point are numerically investigated, the structures of non-spiral streamline pattern are specially analyzed, and the choice of the second direction after the velocity direction is discussed. A brief concluding remarks are given in the last section.

## 2 Swirling degree of velocity field at a point

### 2.1 Definition of swirling degree

Suppose that at one time a fixed region of space $\mathcal{G}$, which may be finite or infinite in extent, is occupied by a continuously distributed fluid; and further, we suppose that the velocity field of the fluid

$$\boldsymbol{v} = V\boldsymbol{n}_1, \tag{1}$$

is a continuously differentiable vector field with zero value at least at one internal point. The points with zero speed in a flow are called the *critical points*. No loss of generality, the Cartesian rectangular coordinate system $(O; x_1, x_2, x_3)$ is used, and the Einstein summation convention for repeated indices, say $\boldsymbol{v} = v_i\mathbf{e}_i$, is adopted.

Now, let us define the **swirling degree** of velocity field at an internal point $p$ as follows. We choose a plane passing through the point $p(\boldsymbol{x})$ and with the normal $\mathbf{m}$ parameterized by

$$\boldsymbol{m} = \boldsymbol{e}_1 \cos\varphi \sin\theta + \boldsymbol{e}_2 \sin\varphi \sin\theta + \boldsymbol{e}_3 \cos\theta, \tag{2}$$

and a convex loop $\mathcal{L}$ on the plane with $\boldsymbol{x}_0 = x_i^0 \boldsymbol{e}_i$ as its center and $\mathcal{D}$ as the domain confined by $\mathcal{L} = \partial\mathcal{D}$. Then an orthonormal base $\{\boldsymbol{m}_1, \boldsymbol{m}_2\}$ of the plane is chosen as

$$\left.\begin{array}{ll} \boldsymbol{m}_1 = \boldsymbol{e}_1, \quad \boldsymbol{m}_2 = \boldsymbol{e}_2, & \text{if } \boldsymbol{m} = \boldsymbol{e}_3; \\ \boldsymbol{m}_1 = \boldsymbol{e}_1 \sin\varphi - \boldsymbol{e}_2 \cos\varphi, \quad \boldsymbol{m}_2 = \boldsymbol{e}_1 \cos\varphi \cos\theta + \boldsymbol{e}_2 \sin\varphi \cos\theta - \boldsymbol{e}_3 \sin\theta, & \text{if } \boldsymbol{m} \neq \boldsymbol{e}_3. \end{array}\right\} \tag{3}$$

So that $\{\boldsymbol{m}, \boldsymbol{m}_1, \boldsymbol{m}_2\}$ constitutes a right-handed frame satisfying $\boldsymbol{m}_1 \times \boldsymbol{m}_2 = \boldsymbol{m}$ and the points on the circular loop centred at $\boldsymbol{x}_0$ have description

$$\boldsymbol{x} = \boldsymbol{x}_0 + \rho(\boldsymbol{m}_1 \cos\phi + \boldsymbol{m}_2 \sin\phi) \text{ or } x_i = x_i^0 + \rho(m_i^1 \cos\phi + m_i^2 \sin\phi). \tag{4}$$

The velocity field on $\mathcal{L}$ is called non-degenerate if $|\boldsymbol{v}(\boldsymbol{y})| > 0$ for all $\boldsymbol{y} \in \mathcal{L}$. For a non-degenerate velocity field on the loop, we obtain a unit vector

$$\boldsymbol{n}_1(\boldsymbol{y}) = \frac{\boldsymbol{v}(\boldsymbol{y})}{|\boldsymbol{v}(\boldsymbol{y})|}, \qquad \boldsymbol{y} \in \mathcal{L}, \tag{5}$$

mapping the point $p(\boldsymbol{y}) \in \mathcal{L}$ onto a point $P(\boldsymbol{n})$ on the unit sphere. When the point $p(\boldsymbol{y})$ moves one circle around the loop $\mathcal{L}$, its mapping onto the unit sphere also goes through a closed curve $\ell$, and so $\mathcal{D}$ onto $\mathscr{d}$. The closed curve on the sphere is assumed to have a unique tangent everywhere so that it is a single or multiple simple closed curve, namely

$$f: \mathcal{L} \to \ell, \tag{6.1}$$

is injective, while

$$f^{-1}: \ell \to \mathcal{L} \tag{6.2}$$

would be multivalued. The smaller area $\mathcal{S}_d$ cut off by the simple closed curve $\ell$ must be no greater than $2\pi$, and is called the **swirling degree** of velocity field on $\mathcal{L}$. Using the unit sphere map (5), Li and Qian (1982) defined the winding number based on the mapping degree. The winding number must be an integer and provides an equivalent description of the swirling degree in 2D space.

For a non-degenerate velocity field on $\mathcal{L}$, the velocity direction $\boldsymbol{n}_1(\boldsymbol{x})$ is mapped to a unique point on the unit sphere. The close curve formed by the velocity field on $\ell$ bounds a curved surface whose area can be approximated by a spherical polygon with vertices $\boldsymbol{q}_I = \boldsymbol{n}_1(\boldsymbol{x}_I)$, where $\boldsymbol{x}_I, I = 1,2,\cdots,N$ are consecutive points on $\mathcal{L}$. Assume that the velocity directions $\boldsymbol{q}_1, \boldsymbol{q}_2, \cdots, \boldsymbol{q}_N = \boldsymbol{q}_0, \boldsymbol{q}_{N+1} = \boldsymbol{q}_1$ correspond to the points $A_1, A_2, \cdots, A_N = A_0, A_{N+1} = A_1$ belong to $\ell$ on the unit sphere, respectively, with adjacent points connected by large arc, and the arc angle between adjacent large arcs $\widehat{A_{k-1}A_k}, \widehat{A_k A_{k+1}}$ is denoted by $\alpha_k = \angle\left(\widehat{A_{k-1}A_k}, \widehat{A_k A_{k+1}}\right) \in [-\pi, \pi)$. Then we have

$$\cos\alpha_k = \frac{\boldsymbol{q}_{k-1} \times \boldsymbol{q}_k}{|\boldsymbol{q}_{k-1} \times \boldsymbol{q}_k|} \cdot \frac{\boldsymbol{q}_k \times \boldsymbol{q}_{k+1}}{|\boldsymbol{q}_k \times \boldsymbol{q}_{k+1}|}, \sin\alpha_k = \left[\left(\frac{\boldsymbol{q}_{k-1} \times \boldsymbol{q}_k}{|\boldsymbol{q}_{k-1} \times \boldsymbol{q}_k|} \times \frac{\boldsymbol{q}_k \times \boldsymbol{q}_{k+1}}{|\boldsymbol{q}_k \times \boldsymbol{q}_{k+1}|}\right) \cdot \boldsymbol{q}_k\right], \tag{7}$$

and the area of the spherical polygon is calculated by (Polyanin and Manzhirov, 2007)

$$\mathcal{S}_d^N = \sum_{k=1}^{N} (\pi - \alpha_k) - (N-2)\pi = 2\pi - \sum_{k=1}^{N} \alpha_k \,. \tag{8}$$

### 2.2 Some formulae for linear velocity field

In this paper, we focus on the linear velocity field with the origin as its critical point given by

$$\boldsymbol{v} = \left(\vartheta - \tfrac{1}{2}\lambda_3\right)\boldsymbol{x} + \tfrac{3}{2}\lambda_3 x_3 \boldsymbol{e}_3 + (\tau_1 x_3 + \sigma\psi x_2)\boldsymbol{e}_1 + [\tau_2 x_3 + (\psi + \tau_3)x_1]\boldsymbol{e}_2 = x_k D_{ki} \boldsymbol{e}_i, \tag{9}$$

with velocity gradient defined by

$$\boldsymbol{D} = \nabla\boldsymbol{v} = \begin{pmatrix} \vartheta - \frac{1}{2}\lambda_3 & \psi + \tau_3 & 0 \\ \sigma\psi & \vartheta - \frac{1}{2}\lambda_3 & 0 \\ \tau_1 & \tau_2 & \vartheta + \lambda_3 \end{pmatrix} \tag{10}$$

under the orthonormal frame determined from the right-eigen representation (Zou *et al*. 2021). In the above, $\sigma = 1$ for the velocity gradient with three real eigenvalues whilst $\sigma = -1$ for the velocity gradient with only one real eigenvalue; $\vartheta$ indicates the compressibility of the flow and $\omega \equiv \sqrt{\psi(\psi + \tau_3)}$ for $\sigma = -1$ indicates the rotation speed of streamline around the extension axis (Zou *et al*. 2021)

$$\boldsymbol{N}_3 = \boldsymbol{e}_3 + c_1\boldsymbol{e}_1 + c_2\boldsymbol{e}_2, \qquad c_1 = \frac{\frac{3}{2}\lambda_3\tau_1 - \psi\tau_2}{\frac{9}{4}\lambda_3^2 + \omega^2}, c_2 = \frac{\frac{3}{2}\lambda_3\tau_2 + \frac{\omega^2}{\psi}\tau_1}{\frac{9}{4}\lambda_3^2 + \omega^2}. \tag{11}$$

The parameter set $\{\vartheta, \lambda_3, \psi, \tau_1, \tau_2, \tau_3\}$ is a good choice in describing the geometric features of streamline pattern.

Using the expression (9) and the definitions of 1-forms

$$\mu_i \equiv \epsilon_{ijk}x_j dx_k = \rho\epsilon_{ijk}x_j^0(m_k^2\cos\phi - m_k^1\sin\phi)d\phi + \rho^2 m_i d\phi, i = 1,2,3, \tag{12}$$

where $\epsilon_{ijk}$ is the permutation symbol, use is made of the Einstein summation convention for repeated indices, and a second-order tensor $\boldsymbol{K}$ defined by

$$K_{ir} = \tfrac{1}{2}\epsilon_{ijk}\epsilon_{rpq}D_{pj}D_{qk} \Leftrightarrow K_{ir}\epsilon_{rpq} = \epsilon_{ijk}D_{pj}D_{qk}, \tag{13.1}$$

which consists of the algebraic cofactors of velocity gradient matrix and satisfies

$$K_{ir}D_{rl} = \tfrac{1}{2}\epsilon_{ijk}\epsilon_{rpq}D_{pj}D_{qk}D_{rl} = \tfrac{1}{2}\det(\boldsymbol{D})\epsilon_{ijk}\epsilon_{ljk} = \delta_{il}\det(\boldsymbol{D}), \tag{13.2}$$

we can derive

$$\boldsymbol{v} \times d\boldsymbol{v} = \boldsymbol{e}_i\epsilon_{ijk}D_{pj}D_{qk}x_p dx_q = \boldsymbol{e}_i K_{ir}\epsilon_{rpq}x_p dx_q = \boldsymbol{e}_i K_{ir}\mu_r. \tag{14}$$

If $\boldsymbol{x}_0$ is the critical point of the velocity field, namely $\boldsymbol{v}_0 = \boldsymbol{e}_p D_{jp}x_j^0 = \boldsymbol{0}$, which results in $\boldsymbol{e}_i K_{ir}\epsilon_{rjk}x_j^0 = \boldsymbol{e}_i\epsilon_{ipq}D_{jp}D_{kq}x_j^0 = \boldsymbol{0}$, substituting (12) into (14) yields

$$\boldsymbol{v} \times d\boldsymbol{v} = \boldsymbol{k}\rho^2 d\phi, \qquad \boldsymbol{k} = \boldsymbol{e}_i K_{ir}m_r, \tag{15.1}$$

or in detail

$$\begin{aligned}\boldsymbol{k} = \boldsymbol{e}_3&[(D_{11}D_{22} - D_{12}D_{21})m_3 + (D_{31}D_{12} - D_{32}D_{11})m_2 + (D_{21}D_{32} - D_{22}D_{31})m_1] \\ +\boldsymbol{e}_2&[(D_{13}D_{21} - D_{11}D_{23})m_3 + (D_{33}D_{11} - D_{31}D_{13})m_2 + (D_{23}D_{31} - D_{21}D_{33})m_1] \\ +\boldsymbol{e}_1&[(D_{12}D_{23} - D_{13}D_{22})m_3 + (D_{32}D_{13} - D_{33}D_{12})m_2 + (D_{22}D_{33} - D_{23}D_{32})m_1].\end{aligned} \tag{15.2}$$

In the frame of right eigen-representation of velocity gradient, $\boldsymbol{k}$ becomes

$$\boldsymbol{k} = (\boldsymbol{e}_1 \quad \boldsymbol{e}_2 \quad \boldsymbol{e}_3)\begin{pmatrix} (\vartheta + \lambda_3)\left(\vartheta - \frac{1}{2}\lambda_3\right) & -(\psi + \tau_3)(\vartheta + \lambda_3) & 0 \\ -\sigma\psi(\vartheta + \lambda_3) & (\vartheta + \lambda_3)\left(\vartheta - \frac{1}{2}\lambda_3\right) & 0 \\ \sigma\tau_2\psi - \tau_1\left(\vartheta - \frac{1}{2}\lambda_3\right) & \tau_1(\psi + \tau_3) - \tau_2\left(\vartheta - \frac{1}{2}\lambda_3\right) & \left(\vartheta - \frac{1}{2}\lambda_3\right)^2 - \sigma\omega^2 \end{pmatrix}\begin{pmatrix} m_1 \\ m_2 \\ m_3 \end{pmatrix}. \tag{15.3}$$

It is easy to see

$$\boldsymbol{m} = \boldsymbol{e}_3 \Rightarrow \boldsymbol{k} = \left[\left(\vartheta - \tfrac{1}{2}\lambda_3\right)^2 - \sigma\omega^2\right]\boldsymbol{e}_3. \tag{16}$$

## 3 Swirl field and vortex classification

### 3.1 Fluid element and formulation of swirl field

The macroscopic velocity characterizing fluid transport must be a statistical average property of a large number of neighboring molecules. Such a molecular cluster has mesoscale and usually remains as a whole during some micro period. A representative of this molecular cluster is unique at every space-time point, called *the fluid element*, and finite fluid can be covered by a finite number of fluid elements. Therefore, the scale of fluid element is small but finite and, two neighboring elements may overlap to each other. The essence of flow is that in addition to the overall migration, fluid elements will form molecular scale stratification and slip over each other. The average slip orientation of fluid element in motion is an internal structure and, requires a kind of isomorphic relation between adjacent fluid elements, called *the swirl field*. The swirl field is sensitive to curved flows: the swirl field vanishes in straight flows, and becomes non-zero in curved flows. Some fluid elements have internal layered structure but may be isotropic; so due to the selectivity of the observer, the swirl field becomes first in dynamics. In other words, the isomorphic structures should be taken account of when formulating the inhomogeneity of velocity field, and affect the viscous interaction between molecules naturally as bending effect of flowing fluid.

Suppose $A_k^i$ denotes the required isomorphic rotation around the $\mathbf{e}_i$ axis when for comparison moving back the fluid element unit distance ahead in the $x_k$ axis direction, then the velocity differential considering the slip isomorphism is denoted as

$$D\boldsymbol{v} = d\boldsymbol{v} + \left(\mathbf{e}_l A_k^l dx_k\right) \times \boldsymbol{v} \equiv (Dv_i)\boldsymbol{e}_i \ \text{ or } \ Dv_i = dv_i + \epsilon_{ilm} A_k^l v_m dx_k. \tag{17}$$

In our previous paper (Zou *et al*. 2021), real vortices are divided into three categories: small-scale eddies, stable vortices which are able to be observed through streamline but may be missed by a moving observer, and moving vortices carried and distorted by the turbulent mainstream. The statistical description of small-scale eddies among a fluid element is called *the micro-eddy field* $\Phi^i$, which is suitable to combined with the swirl field to form an axial-vector-valued differential 1-form

$$\mathbf{W}^i = A_k^i dx_k + \Phi^i dt, \tag{18}$$

indicating the spatiotemporal structure in the fluid element, and wholly called *the vortex field*. This new field is in general independent of the velocity field (Fig. 2). The dynamics of vortex field coupling with the velocity field is a story too big to be discussed in this paper.

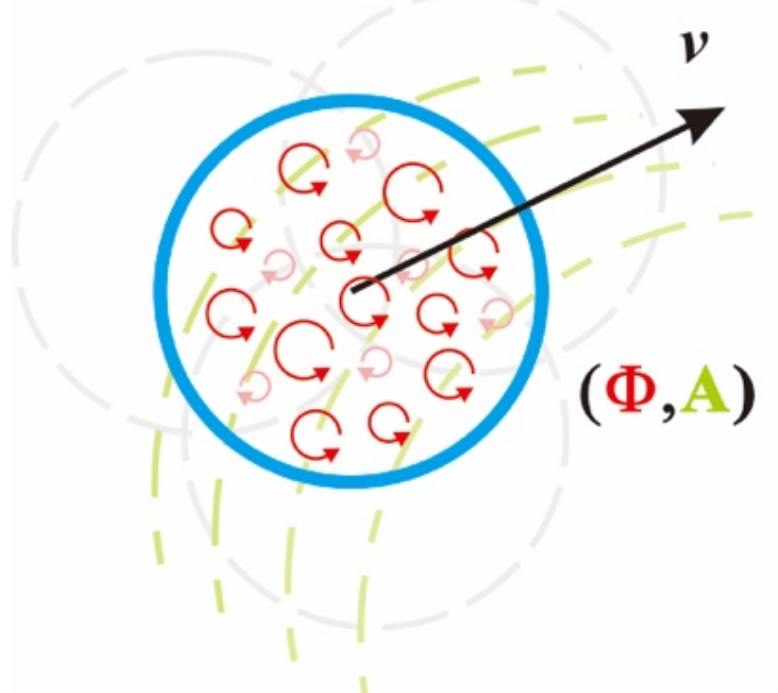

**Figure 2.** Fluid element model

Now we focus on the steady flows with stable vortices, especially the simplest inhomogeneous flow described by linear velocity field. In steady flows, the streamlines are fixed and define the way of fluid translating from one point to another, since the fluid on both sides of a streamline has no exchange other than microscopic diffusion. The determination of another direction to complete the contact relation comes from the coupling mechanism of streamwise vortex. Here we use the streamline bending, namely $\mathbf{n}_1 \cdot \nabla \mathbf{n}_1$ to introduce the second direction $\mathbf{n}_2$, as shown in Fig. 3. More consideration is left to the discussion in Section 5.3. Two sets of curves are extended point by point along two mutually orthogonal directions, but may not be woven into layered surfaces. The third direction $\mathbf{n}_3$ is constructed to form a right-hand orthogonal frame. All fluid elements with definite frames $(\mathbf{n}_1, \mathbf{n}_2, \mathbf{n}_3)$ defines a regular region of fluid, whilst the isotropic region is trivial with fluid in static or straight flow. What we are concerned about is the structures of curved flow around the critical points (lines).

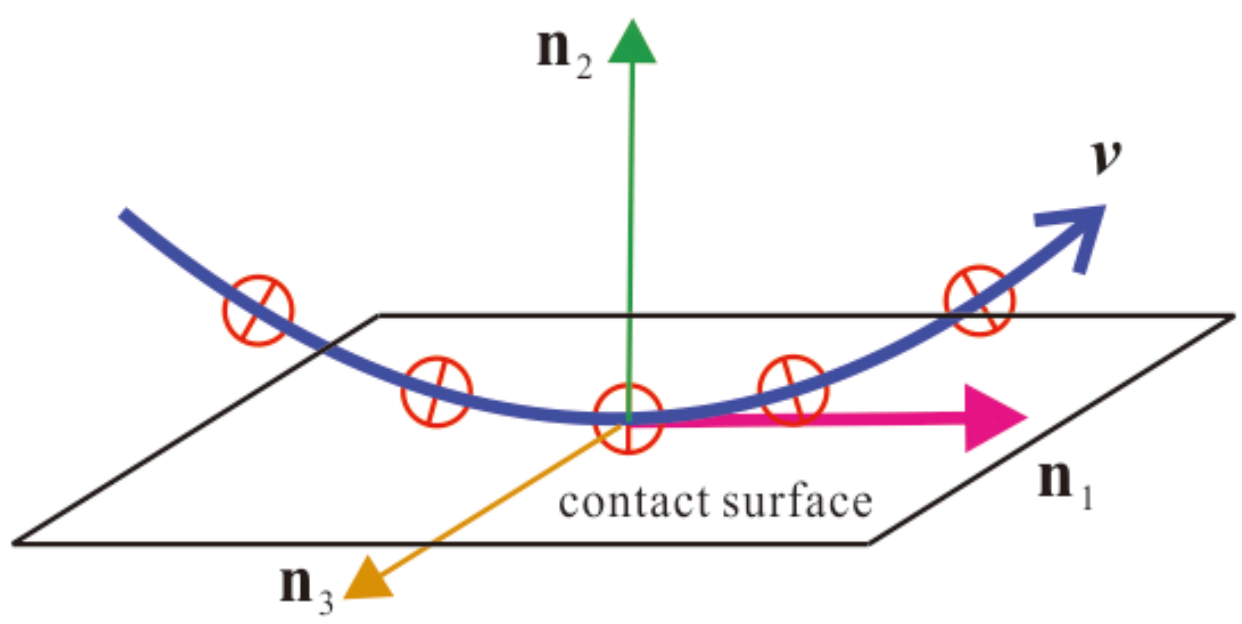

**Figure 3.** Curved streamline and contact surface

In the regular region of flow, the set of Cartesian bases at any point $p$ is assumed to be $(p; \mathbf{e}_1, \mathbf{e}_2, \mathbf{e}_3)$, and the contact orientation frame uniquely coming from the velocity field (1) is denoted by

$$\mathbf{n}_i = R_{ij}\mathbf{e}_j. \tag{19}$$

Thus, the corresponding coframe $(p; \omega_1, \omega_2, \omega_3)$ (Chern, Chen and Lam, 1999) satisfies the inner product relations: $\langle \mathbf{n}_i, \omega_j \rangle = \delta_{ij}$, i.e.

$$\omega_i = R_{ij} dx_j. \tag{20}$$

Using the new frame, the differential of position vector at the regular point $p$ can be written as

$$d\mathbf{r} = \mathbf{e}_i dx_i = \mathbf{n}_i \omega_i, \tag{21}$$

and further the differential of frame is given by

$$d\mathbf{n}_i = \omega_{ij}\mathbf{n}_j, \tag{22}$$

satisfying $\omega_{ij} + \omega_{ji} = 0$. Thereby, we can integrate the streamline and slip line tangent to $\mathbf{n}_1$ and $\mathbf{n}_2$ everywhere, and the corresponding arc length coordinates are denoted by $s_1$ and $s_2$, respectively. When the rotation transformation $R_{ij}(\mathbf{r})$ is obtained, the differential 1-forms $\omega_i$ are calculated by (20)2, and the connection 1-forms $\omega_{ij}$ can be expressed by

$$\omega_{ij} = (dR_{ik})R_{jk} = (\partial_m R_{ik})R_{jk}dx_m = (\partial_m R_{ik})R_{jk}R_{nm}\omega_n. \tag{23}$$

The following analyses of the slip structure are based on this pair of frame fields.

### 3.2 Slip isomorphism and direct calculation formula

The orientation isomorphism between adjacent fluid elements stems from their overlap so that their velocities should be compared under the same contact frame, or equivalently in so-called parallel translation, a micro-rotation should be carried out according to the difference of the contact frames. Such a connection structure is characterized by an axial-vector-valued spatial 1-form, that

$$D\mathbf{e}_i = -\epsilon_{ilm}A_k^l \mathbf{e}_m dx_k = \left(A_k^l \mathbf{e}_l dx_k\right) \times \mathbf{e}_i. \tag{24}$$

For the case that the contact frame is completely determined by the velocity direction, say (23), the orientation isomorphism means

$$0 = D\mathbf{n}_i = d\mathbf{n}_i + \left(A_k^l \mathbf{e}_l dx_k\right) \times \mathbf{n}_i = \omega_{ij}\mathbf{n}_j + \left(A_k^l \mathbf{e}_l dx_k\right) \times \mathbf{n}_i, \tag{25}$$

yielding the relation

$$A_k^i dx_k = -\tfrac{1}{2}R_{pi}\epsilon_{plm}\omega_{lm} = \tfrac{1}{2}\epsilon_{ilm}\left(dR_{pl}\right)R_{pm}. \tag{26}$$

In region with differentiable orientation field $R_{ij}$, the expression (26) implies an integrable swirl field satisfying

$$\mathbf{B}^i = d\mathbf{A}^i + \tfrac{1}{2}\epsilon_{ilm}\mathbf{A}^l \wedge \mathbf{A}^m \equiv 0. \tag{27}$$

In order to obtain the direct formula of the swirl field in terms of the velocity, we start from (26)1 and divide it into two parts: the one perpendicular to the velocity direction and the one parallel to the velocity direction; that is

$$A_k^i dx_k = -R_{1i}\omega_{23} - (R_{2i}\omega_{31} + R_{3i}\omega_{12}), \tag{28.1}$$

where the asymmetric property of $\omega_{ij}$ is made use of. From (23)1 we have

$$\omega_{23} = (dR_{2k})R_{3k} = \mathbf{n}_3 \cdot d\mathbf{n}_2 = \tfrac{1}{2}(\mathbf{n}_3 \cdot d\mathbf{n}_2 - \mathbf{n}_2 \cdot d\mathbf{n}_3) \tag{28.2}$$

and

$$R_{2i}\omega_{32} + R_{3i}\omega_{12} = -R_{2i}R_{3k}dR_{1k} + R_{3i}R_{2k}dR_{1k} = \epsilon_{1pq}R_{pk}R_{qi}dR_{1k} = \epsilon_{imk}R_{1m}dR_{1k}. \tag{28.3}$$

Substituting them into (28.1) yields

$$A_k^l \mathbf{e}_l dx_k = -\boldsymbol{n}_1 \times d\boldsymbol{n}_1 + \tfrac{1}{2}(\mathbf{n}_2 \cdot d\mathbf{n}_3 - \mathbf{n}_3 \cdot d\mathbf{n}_2)\boldsymbol{n}_1. \tag{28.4}$$

Considering the coupling (17) with the velocity field, it is easy to prove only the first part of the swirl field (28.4) has contribution to the covariant gradient of the velocity.

## 4 Streamline pattern and swirl field

### 4.1 Linear velocity fields in 2D space

The streamline pattern of 2D linear velocity field

$$\boldsymbol{v} = [\vartheta x + (\psi + \tau)y]\mathbf{e}_1 + (\sigma\psi x + \vartheta y)\mathbf{e}_2 \tag{29}$$

can be classified into six categories (Zou *et al.*, 2021), here condensed into four types, as shown in Table 1. Direct investigation shows that the swirling degree gets three values $2\pi, 0, -2\pi$, and the criteria

$$\Delta = -\sigma\psi(\psi + \tau) = -\sigma\omega^2 \tag{30}$$

is not a topological parameter. Mathematically, the swirling degree in 2D space can be judged by the sign of $\det(\boldsymbol{D})$, called the index of linear velocity field (Outerelo and Ruiz, 2009).

**Table 1**. Topological index of streamline patterns and its identification

| **Name** | **Parameter** | **Swirling degree** | **Loop integration of** $A_k^3 dx_k$ |
|---|---|---|---|
| Spiral flow | $\Delta > 0$ | $2\pi\ (C); 0\ (NC)$ | $-2\pi\ (C); 0\ (NC)$ |
| Nodal flow | $\Delta \leq 0, \lvert \vartheta \rvert > \beta$ | $2\pi\ (C); 0\ (NC)$ | $-2\pi\ (C); 0\ (NC)$ |
| Straight flow | $\Delta \leq 0, \lvert \vartheta \rvert = \beta$ | - | 0 |
| Saddle flow | $\Delta < 0, \lvert \vartheta \rvert < \beta$ | $-2\pi\ (C); 0\ (NC)$ | $2\pi\ (C); 0\ (NC)$ |

**Note**: "C" indicates the integration domain containing the critical point; and "NC" not.

In 2D flows, the swirl field has fixed rotation direction, can be expressed from (32) as

$$A_k^3 dx_k = -\frac{\boldsymbol{v} \times d\boldsymbol{v}}{V^2} \cdot \mathbf{e}_3 = \frac{(\sigma\omega^2 - \vartheta^2)(xdy - ydx)}{[\vartheta x + (\psi + \tau)y]^2 + (\sigma\psi x + \vartheta y)^2}, \tag{31.1}$$

which will reduce in the polar coordinate $(\rho, \theta)$ to

$$A_k^3 dx_k = \frac{(\sigma\omega^2 - \vartheta^2)d\theta}{[\vartheta\cos\theta + (\psi + \tau)\sin\theta]^2 + (\sigma\psi\cos\theta + \vartheta\sin\theta)^2}. \tag{31.2}$$

Making use of the integral formula

$$\int_0^{2\pi} \frac{1}{1 + a\cos x} dx = \frac{2\pi}{\sqrt{1 - a^2}}, |a| < 1, \tag{32}$$

we obtain the integration for $\mathcal{L}$ contains the origin

$$\oint_{\mathcal{L}_C} A_k^3 dx_k = 2\pi \frac{\sigma\omega^2 - \vartheta^2}{|\sigma\omega^2 - \vartheta^2|} = \begin{cases} -2\pi, & \Delta > 0; \\ -2\pi, & \Delta < 0, |\vartheta| > \omega; \\ +2\pi, & \Delta < 0, |\vartheta| < \omega. \end{cases} \tag{33.1}$$

For the case $V \neq 0$ in $\mathcal{D}$, due to

$$d(A_k^3 dx_k) = \frac{2(\sigma\omega^2 - \vartheta^2)dx \wedge dy}{[\vartheta x + (\psi + \tau)y]^2 + (\sigma\psi x + \vartheta y)^2} - \frac{2(\sigma\omega^2 - \vartheta^2)dx \wedge dy}{[\vartheta x + (\psi + \tau)y]^2 + (\sigma\psi x + \vartheta y)^2} \equiv 0,$$

we always have

$$\oint_{\mathcal{L}_{NC}} A_k^3 dx_k \equiv 0. \tag{33.2}$$

Thus, the loop integral property of the swirl field completely coincides with the swirling degree (a minus sign difference). Can we say the swirl field is a topological field of contact relation in fluid?

### 4.2 Linear velocity fields in 3D space

#### 4.2.1 Formula of the swirling degree

The swirling degree in 3D flows is not so intuitive, and looks like confusing. In Fig. 4, the loop $\mathcal{L}$ with normal $\mathbf{m} = \mathbf{e}_3$ and radius 2 at the point $(0, 0, 3)$ and its unit sphere map are shown. Calculations indicate that the area on the unit sphere depends the point $\boldsymbol{x}_0$, the normal $\boldsymbol{m}$ and the loop $\mathcal{L}$, even though the maps of $\mathcal{L}_O$ and $\mathcal{L}_{NO}$ are quite different.

In Fig. 4, the $x_3$-axis is the right eigendirection of the real eigenvalue $\vartheta + \lambda_3$, while

$$\boldsymbol{N}_3 = \mathbf{e}_3 + c_2\mathbf{e}_2 + c_1\mathbf{e}_1, \tag{34.1}$$

is the corresponding left eigenvector with the coefficients $c_1, c_2$ defined by [Zou *et al.* (2021)]

$$c_1 = \frac{\frac{3}{2}\lambda_3\tau_1 + \sigma\psi\tau_2}{\frac{9}{4}\lambda_3^2 - \sigma\omega^2}, c_2 = \frac{\frac{3}{2}\lambda_3\tau_2 + (\psi + \tau_3)\tau_1}{\frac{9}{4}\lambda_3^2 - \sigma\omega^2}. \tag{34.2}$$

In order to obtain the general formula of the swirling degree, we introduce the relation

$$\mathbf{n}_1 = \boldsymbol{e}_1\cos\alpha\cos\beta + \boldsymbol{e}_2\sin\alpha\cos\beta + \boldsymbol{e}_3\sin\beta, \qquad -\pi/2 \leq \beta \leq \pi/2, 0 \leq \alpha < 2\pi. \tag{35}$$

Then, the mapping area can be calculated by

$$\mathcal{S}_d = \iint_d \cos\beta d\alpha \wedge d\beta \tag{36}$$

if the domain $d$ is known. On the other hand, from (29) we extract the part of the swirl field perpendicular to the velocity direction, that is (called *the normal part* in the following)

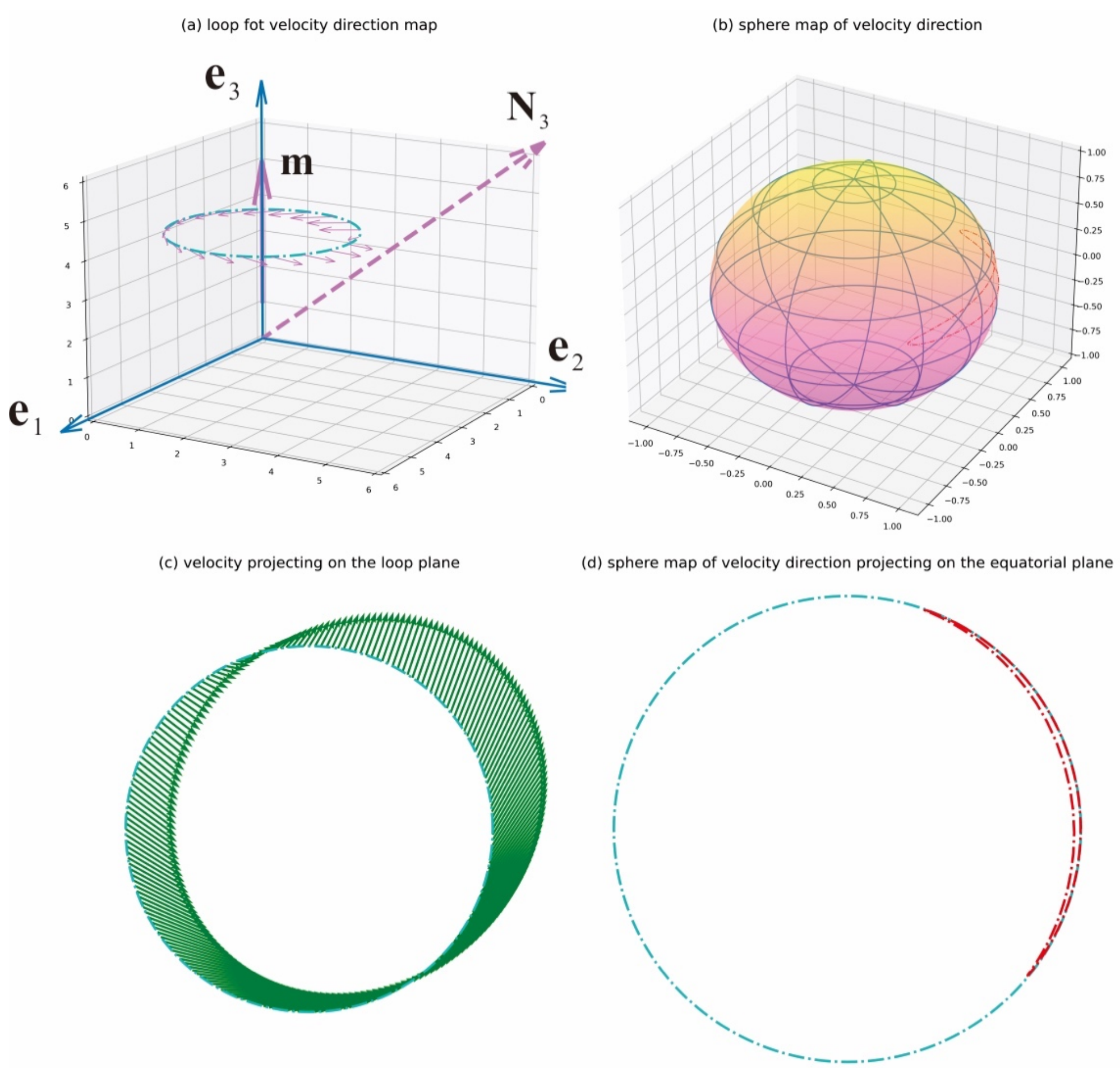

**Figure 4**. Loop and mapping for the swirling degree (complex eigenvalues case): (a) loop for velocity direction map, (b) sphere map of velocity direction, (c) velocity projecting on the loop plane, (d) sphere map of velocity direction projecting on the equatorial plane.

$$\hat{A}_k^l \mathbf{e}_l dx_k = -\frac{\boldsymbol{v}\times d\boldsymbol{v}}{V^2} = -\mathbf{n}_1 \times d\mathbf{n}_1, \tag{37}$$

and derive the loop integration as

$$\mathrm{Int}_{\mathcal{L}}\, \hat{\boldsymbol{A}} \triangleq \oint_{\mathcal{L}} \hat{A}_k^l \mathbf{e}_l dx_k = \iint_{\mathcal{D}} d\left(\hat{A}_k^l \mathbf{e}_l dx_k\right) = -\iint_d d(\mathbf{n}_1 \times d\mathbf{n}_1) = -2\iint_d \mathbf{n}_1 \cos\beta d\alpha \wedge d\beta\,. \tag{38}$$

It should be noticed the integral derivation works only for the loop without any critical point in $\mathcal{D}$ and $\partial\mathcal{D}$. Therefore, by comparing (36) and (38), we can conclude two important relations:

$$\mathcal{S}_d = -\frac{1}{2}\iint_{\mathcal{D}} \mathbf{n}_1 \cdot d\left(\hat{A}_k^l \mathbf{e}_l dx_k\right), \tag{39}$$

$$\left|\mathrm{Int}_{\mathcal{L}}\, \hat{\boldsymbol{A}}\right| = 2\left|\iint_d \mathbf{n}_1 \cos\beta d\alpha \wedge d\beta\right| \le 2\left|\iint_d \cos\beta d\alpha \wedge d\beta\right| = 2\mathcal{S}_d. \tag{40}$$

Another interesting deduction is: $\mathcal{S}_d = 0 \Rightarrow d = \emptyset \Rightarrow \mathrm{Int}_{\mathcal{L}}\, \hat{\boldsymbol{A}} = \mathbf{0}$.

### 4.2.2 Integration in $\mathcal{D}$ without any critical point

The following derivation is universal

$$d\left(\frac{\boldsymbol{v}\times d\boldsymbol{v}}{V^2}\right)_i = \epsilon_{ilm}\frac{dv_l\wedge dv_m}{V^2} - 2v_l dv_l \wedge \frac{\epsilon_{ikm}v_k dv_m}{V^4} = \epsilon_{ilm}\frac{dv_l\wedge dv_m}{V^2} - \epsilon_{rlm}\epsilon_{rpq}v_l dv_p\wedge\frac{\epsilon_{ikm}v_k dv_q}{V^4}$$

$$= \epsilon_{ilm}\frac{dv_l\wedge dv_m}{V^2} - \epsilon_{rpq}(v_kv_k\delta_{ir} - v_iv_r)\frac{dv_p\wedge dv_q}{V^4} = \frac{v_i}{V^4}\epsilon_{rpq}v_r dv_p\wedge dv_q. \tag{41.1}$$

For a linear velocity field with the origin as its critical point, we have $v_i = x_k D_{ki}$ from (10), and further

$$d\left(\frac{\boldsymbol{v}\times d\boldsymbol{v}}{V^2}\right)_i = \frac{v_i}{V^4}\epsilon_{rpq}x_l D_{lr}D_{mp}D_{nq}dx_m\wedge dx_n = \det(\boldsymbol{D})\frac{v_i}{V^4}\epsilon_{lmn}x_l dx_m\wedge dx_n = 2\det(\boldsymbol{D})\frac{v_i}{V^4}x_l da_l, \tag{41.2}$$

where use is made of the area element

$$da_k = \tfrac{1}{2}\epsilon_{ijk}dx_i\wedge dx_j. \tag{41.3}$$

For the case $\det(\boldsymbol{D}) = (\vartheta + \lambda_3)\left[\left(\vartheta - \tfrac{1}{2}\lambda_3\right)^2 - \sigma\omega^2\right] = 0$, that means $\vartheta + \lambda_3 = 0$, and/or $\left(\vartheta - \tfrac{1}{2}\lambda_3\right)^2 - \omega^2 = 0$ when $\sigma = 1$. In any case, we can find a left eigendirection with zero eigenvalue, and the flow is confirmed in planes with the corresponding right eigendirection as their normal. No loss of generality, we take $\vartheta + \lambda_3 = 0$ as the result of $\det(\boldsymbol{D}) = 0$. Substitution of (4) yields

$$d\left(\frac{\boldsymbol{v}\times d\boldsymbol{v}}{V^2}\right) = 2\det(\boldsymbol{D})\frac{\boldsymbol{v}}{V^4}x_m da_m = 2x_k^0 m_k \det(\boldsymbol{D})\frac{\boldsymbol{v}}{V^4}\rho d\rho\wedge d\phi, \tag{42}$$

where uses are made of

$$da_p = \tfrac{1}{2}\epsilon_{pij}dx_i\wedge dx_j = m_p\rho d\rho\wedge d\phi, \qquad x_k m_k = x_k^0 m_k. \tag{43}$$

According to the definition (40) and the relation (42), we use the Stokes integral theorem under the condition that $V \neq 0$ on $\mathcal{D}$ is satisfied to obtain the operable formulae

$$\mathrm{Int}_{\mathcal{L}_{NC}}\widehat{\boldsymbol{A}} = -\oint_{\mathcal{L}_{NC}}\frac{\boldsymbol{v}\times d\boldsymbol{v}}{V^2} = -\iint_{\mathcal{D}} d\left(\frac{\boldsymbol{v}\times d\boldsymbol{v}}{V^2}\right) = -2x_k^0 m_k\det(\boldsymbol{D})\iint_{\mathcal{D}}\frac{\boldsymbol{v}}{V^4}\rho d\rho\wedge d\phi\,, \tag{44}$$

$$\mathcal{S}_d = x_k^0 m_k \det(\boldsymbol{D})\iint_{\mathcal{D}}\frac{1}{V^3}\rho d\rho\wedge d\phi\,. \tag{45}$$

The direct deductions are ($\boldsymbol{x}_0 = \boldsymbol{0}$ is impossible in this situation)

$$\mathrm{Int}_{\mathcal{L}_{NC}}\widehat{\boldsymbol{A}} = \boldsymbol{0}, \mathcal{S}_d = 0 \quad \text{if } x_k^0 m_k = 0 \text{ or } \det(\boldsymbol{D}) = 0. \tag{46}$$

The equivalence between formulae (8) and (45), $(38)_1$ and $(44)_3$, and the results (46) are verified numerically.

The triviality condition $x_k^0 m_k = 0$ indicates that the plane of the loop $\mathcal{L}_{NC}$ passes through the origin (critical point). The condition $\det(\boldsymbol{D}) = 0$ implies that $\boldsymbol{N}_3 \times \boldsymbol{x} = \boldsymbol{0}$ defines a critical line, and as a triviality condition tells all swirling degrees of $\mathcal{L}_{NC}$ vanish when a critical straight line exists.

### 4.2.3 Integration with $\boldsymbol{x}_0$ as an isolated critical point

When $\boldsymbol{x}_0$ is the only critical point in $\mathcal{D}$, we have $\boldsymbol{v}(\boldsymbol{x}_0) = \boldsymbol{0}$ and make use of the expressions (4) and $(9)_2$ to obtain

$$V^2 = D_{pk}x_p D_{qk}x_q = \rho^2 D_{pk}D_{qk}\left(m_p^1\cos\phi + m_p^2\sin\phi\right)\left(m_q^1\cos\phi + m_q^2\sin\phi\right)$$
$$= \rho^2(a_k\cos\phi + b_k\sin\phi)(a_k\cos\phi + b_k\sin\phi) \tag{47.1}$$

with

$$a_k = m_p^1 D_{pk}, b_k = m_p^2 D_{pk}; \tag{47.2}$$

and further

$$\frac{V^2}{\rho^2} = \frac{a_k a_k + b_k b_k}{2} + \frac{a_k a_k - b_k b_k}{2}\cos 2\phi + a_k b_k \sin 2\phi = \frac{a_k a_k + b_k b_k}{2} + A\cos 2(\phi + \phi_0) \tag{47.3}$$

with

$$A^2 = \left(\frac{a_k a_k - b_k b_k}{2}\right)^2 + (a_k b_k)^2, \qquad \sin 2\phi_0 = \frac{a_k b_k}{A}. \tag{47.4}$$

In virtue of these derivation and the integral formula (32), we have

$$\int_0^{2\pi}\frac{\rho^2}{V^2}d\phi = \frac{2\pi}{|\boldsymbol{k}|}, \tag{47.5}$$

because

$$\left(\frac{a_k a_k + b_k b_k}{2}\right)^2 - A^2 = (a_k a_k)(b_l b_l) - (a_k b_k)^2 = \left(\epsilon_{pij} a_i b_j\right)\left(\epsilon_{pkl} a_k b_l\right)$$
$$= \left(\epsilon_{pij} D_{ri} D_{qj} m_r^1 m_q^2\right)\left(\epsilon_{pkl} D_{sk} D_{tl} m_s^1 m_t^2\right)$$
$$= \left(K_{pj} \epsilon_{jrq} m_r^1 m_q^2\right)\left(K_{pk} \epsilon_{kst} m_s^1 m_t^2\right) = \left(K_{pj} m_j\right)\left(K_{pk} m_k\right) = |\boldsymbol{k}|^2 \tag{47.6}$$

where $(15.1)_2$ and relation $(13.1)_2$ are made use of.

Using (15.1), we finally reach the explicit formula

$$\mathrm{Int}_{\mathcal{L}_C} \widehat{\boldsymbol{A}} = -\oint_{\mathcal{L}_C} \frac{\boldsymbol{v} \times d\boldsymbol{v}}{V^2} = -\int_0^{2\pi} \frac{\rho^2 \boldsymbol{k}}{V^2} d\phi = -2\pi \widehat{\boldsymbol{k}}, \qquad \widehat{\boldsymbol{k}} = \frac{\boldsymbol{k}}{|\boldsymbol{k}|}. \tag{48}$$

The direction $\widehat{\boldsymbol{k}}$ of topological index is called an intrinsic direction if it is the same as the surface normal $\boldsymbol{m}$. From (16), we see that $\boldsymbol{e}_3$ is an intrinsic direction. Further if the real eigenvalue $\vartheta + \lambda_3 = 0$, all points satisfying $\boldsymbol{N}_3 \times \boldsymbol{x} = \boldsymbol{0}$ form a critical line, and (15.3) reduces to

$$\boldsymbol{k} = \boldsymbol{K} \cdot \boldsymbol{m} = \left(\frac{9}{4}\lambda_3^2 - \sigma\omega^2\right)(\boldsymbol{N}_3 \cdot \boldsymbol{m})\boldsymbol{e}_3. \tag{49}$$

Since $\boldsymbol{x}_0$ should be the only critical point in $\mathcal{D}$, the critical straight line cannot be in $\mathcal{D}$, so $\boldsymbol{N}_3 \cdot \boldsymbol{m} \neq 0$, and (48) becomes

$$\mathrm{Int}_{\mathcal{L}_C} \widehat{\boldsymbol{A}} = -2\pi \boldsymbol{e}_3, \tag{50}$$

that means in this case the direction $\widehat{\boldsymbol{k}}$ of topological index is fixed to be $\pm\boldsymbol{e}_3$. About $\mathcal{S}_d$ on $\mathcal{L}_C$, the direct numerical calculation from (8) shows that it always converges to $2\pi$.

#### 4.2.4 Extension and non-commutativity effect of 3D rotation group

At the beginning, the loop is assumed to be a circle centered at $\boldsymbol{x}_0$. But the derivations, say (42) for example, can be extended to a convex domain with $\boldsymbol{x}_0$ as an interior point. Together with the additivity of integration on the non-degenerate loop, the results in Section 4.2.2 can be extended to any simple domain $\mathcal{D}$ in a plane and its boundary $\mathcal{L} = \partial\mathcal{D}$, and the integrations on $\mathcal{L}_C$ in Section 4.2.3 can boil down to the loop of a micro-circle centered at the critical point; that is to say, the results on $\mathcal{L}_C$ is completely the property of the critical point. Three main points for a linear velocity field are summed up as follows:

- $\left|\mathrm{Int}_{\mathcal{L}_{NC}} \widehat{\boldsymbol{A}}\right| = \mathcal{S}_d = 0$ when there is a critical straight line or the plane of $\mathcal{L}_{NC}$ contains the critical point;
- $\left|\mathrm{Int}_{\mathcal{L}_C} \widehat{\boldsymbol{A}}\right| = \mathcal{S}_d = 0$ if the plane of $\mathcal{L}_C$ contains the critical straight line, else $\left|\mathrm{Int}_{\mathcal{L}_C} \widehat{\boldsymbol{A}}\right| = \mathcal{S}_d = 2\pi$;
- The normal part of the swirl field is a more generalized topological quantity related to streamline pattern, and has property: $\left|\mathrm{Int}_{\mathcal{L}} \widehat{\boldsymbol{A}}\right| \leq \min(2\pi, 2\mathcal{S}_d)$.

First two are very similar to the case of 2D flows, but how to understand the nonzero $\mathcal{S}_d$ in Fig. 4?

As mentioned earlier in (44), the loop integration on $\mathcal{L}_{NC}$ can be transformed into the area integration through the Stokes integral theorem. Let us calculate a special term

$$\epsilon_{ilm} \widehat{\mathbf{A}}^l \wedge \widehat{\mathbf{A}}^m = \epsilon_{ilm} \frac{\epsilon_{lpq} v_p dv_q}{V^2} \wedge \frac{\epsilon_{mnr} v_n dv_r}{V^2} = \frac{v_i}{V^4} \epsilon_{lpq} v_p dv_q \wedge dv_l = -d\widehat{\mathbf{A}}^i, \tag{51}$$

it seems that a generalized integration

$$\langle \mathrm{Int}_{\mathcal{L}} \widehat{\boldsymbol{A}}^i \rangle \equiv \oint_{\mathcal{L}} \widehat{\boldsymbol{A}}^i + \iint_{\mathcal{D}} \epsilon_{ilm} \widehat{\mathbf{A}}^l \wedge \widehat{\mathbf{A}}^m = -\oint_{\mathcal{L}} \frac{\boldsymbol{v} \times d\boldsymbol{v}}{V^2} + \iint_{\mathcal{D}} d\left(\frac{\boldsymbol{v} \times d\boldsymbol{v}}{V^2}\right) \tag{52}$$

can be introduced to identify the singularity structure in both 2D and 3D flows, by the reason that the 3D rotations are in general non-commutative. When the flow becomes two-dimensional, the additional part vanishes identically. But it seems to be a trick, as shown in the next section, the distribution of $\mathrm{Int}_{\mathcal{L}} \widehat{\boldsymbol{A}}^i$ is also meaningful.

A deeper consideration is from the integrability of the whole swirl field. For a well-defined contact orientation frame $(\mathbf{n}_1, \mathbf{n}_2, \mathbf{n}_3)$, from (28.4) we have

$$\mathbf{A}^i \mathbf{e}_i = \widehat{\mathbf{A}}^i \mathbf{e}_i + \tfrac{1}{2}(\mathbf{n}_2 \cdot d\mathbf{n}_3 - \mathbf{n}_3 \cdot d\mathbf{n}_2)\boldsymbol{n}_1. \tag{53.1}$$

Since $\boldsymbol{n}_1 \times d\mathbf{n}_1$ is perpendicular to $\boldsymbol{n}_1$, and so $\mathbf{e}_i \epsilon_{ilm} (\boldsymbol{n}_1 \times d\mathbf{n}_1)_l \wedge (\boldsymbol{n}_1 \times d\mathbf{n}_1)_m$ is parallel to $\boldsymbol{n}_1$, the following derivation holds:

$$d\mathbf{n}_2 \dot{\wedge} d\mathbf{n}_3 = (\boldsymbol{n}_1 \cdot d\mathbf{n}_2) \wedge (\boldsymbol{n}_1 \cdot d\mathbf{n}_3) = (\boldsymbol{n}_2 \cdot d\mathbf{n}_1) \wedge (\boldsymbol{n}_3 \cdot d\mathbf{n}_1) = -\boldsymbol{n}_3 \cdot (\boldsymbol{n}_1 \times d\mathbf{n}_1) \wedge \boldsymbol{n}_2 \cdot (\boldsymbol{n}_1 \times d\mathbf{n}_1)$$

$$= \tfrac{1}{2}[\boldsymbol{n}_2 \cdot (\boldsymbol{n}_1 \times d\mathbf{n}_1) \wedge \boldsymbol{n}_3 \cdot (\boldsymbol{n}_1 \times d\mathbf{n}_1) - \boldsymbol{n}_3 \cdot (\boldsymbol{n}_1 \times d\mathbf{n}_1) \wedge \boldsymbol{n}_2 \cdot (\boldsymbol{n}_1 \times d\mathbf{n}_1)]$$

$$\Rightarrow (d\mathbf{n}_2 \dot{\wedge} d\mathbf{n}_3)\boldsymbol{n}_1 = \tfrac{1}{2}\epsilon_{ilm}\mathbf{e}_i(\boldsymbol{n}_1 \times d\mathbf{n}_1)_l \wedge (\boldsymbol{n}_1 \times d\mathbf{n}_1)_m = \tfrac{1}{2}\epsilon_{ilm}\mathbf{e}_i\widehat{\mathbf{A}}^l \wedge \widehat{\mathbf{A}}^m. \tag{53.2}$$

Thus, we have

$$d\mathbf{A}^i = d\widehat{\mathbf{A}}^i + d\boldsymbol{n}_1 \wedge (\mathbf{n}_2 \cdot d\boldsymbol{n}_3) + (d\mathbf{n}_2 \dot{\wedge} d\mathbf{n}_3)\boldsymbol{n}_1 = d\widehat{\mathbf{A}}^i + \tfrac{1}{2}\epsilon_{ilm}\mathbf{e}_i\widehat{\mathbf{A}}^l \wedge \widehat{\mathbf{A}}^m + d\boldsymbol{n}_1 \wedge (\mathbf{n}_2 \cdot d\boldsymbol{n}_3), \tag{53.3}$$

and

$$\epsilon_{ilm}\mathbf{A}^l \wedge \mathbf{A}^m = \epsilon_{ilm}\widehat{\mathbf{A}}^l \wedge \widehat{\mathbf{A}}^m - 2[(\mathbf{n}_1 \times d\boldsymbol{n}_1) \times \mathbf{n}_1] \wedge (\mathbf{n}_2 \cdot d\boldsymbol{n}_3) = \epsilon_{ilm}\widehat{\mathbf{A}}^l \wedge \widehat{\mathbf{A}}^m - 2d\mathbf{n}_1 \wedge (\mathbf{n}_2 \cdot d\boldsymbol{n}_3). \tag{53.4}$$

Combination of (53.3) and (53.4) yields

$$\mathbf{B}^i = d\mathbf{A}^i + \tfrac{1}{2}\epsilon_{ilm}\mathbf{A}^l \wedge \mathbf{A}^m = d\widehat{\mathbf{A}}^i + \epsilon_{ilm}\widehat{\mathbf{A}}^l \wedge \widehat{\mathbf{A}}^m; \tag{53.5}$$

this means that the treatment (52) is a natural expression for the integrability (27) of the swirl field, namely the existence of orientation frame. The above deduction is independent of the choice of $\mathbf{n}_2$ and $\mathbf{n}_3$.

## 5 Numerical analysis and discussion

### 5.1 Dual directivity of $\mathrm{Int}_{\mathcal{L}}\, \widehat{\boldsymbol{A}}^i$ around the isolated critical point

The dual directivity of spiral streamline pattern can be described by the right eigen-representation of velocity gradient, but a critical point is necessary to locate it. The swirling degree is a scalar, the plane $x_3 = 0$ is not special for $\mathcal{S}_d$: for all $\mathcal{D}$-plane containing the origin $O$, $\mathcal{S}_d = 2\pi$ if $O \in \mathcal{D}$; else $\mathcal{S}_d = 0$. But $\mathrm{Int}_{\mathcal{L}}\, \widehat{\boldsymbol{A}}$ can identify the base plane, as shown in (16), for $O \in \mathcal{D}$, only $\boldsymbol{m} = \boldsymbol{e}_3 \Rightarrow \mathrm{Int}_{\mathcal{L}}\, \widehat{\boldsymbol{A}} = \pm 2\pi\boldsymbol{e}_3$, where the minus happens when $\sigma = 1$ and $\left(\vartheta - \frac{1}{2}\lambda_3\right)^2 < \omega^2$. This is one of the meanings of the rotation direction of the spiral streamline.

The radial line along the extension direction $\boldsymbol{N}_3$ can be recognized by both $\mathcal{S}_d$ and $\left|\mathrm{Int}_{\mathcal{L}}\, \widehat{\boldsymbol{A}}\right|$. As shown in Fig. 5 with the velocity gradient

$$\boldsymbol{D} = \begin{bmatrix} -0.008212 & 0.302153 & 0 \\ -0.185413 & -0.008212 & 0 \\ 0.171535 & 0.084078 & 0.014756 \end{bmatrix},$$

their iso-surfaces all clearly catch the extension direction, except that $\mathcal{S}_d$ decays faster.

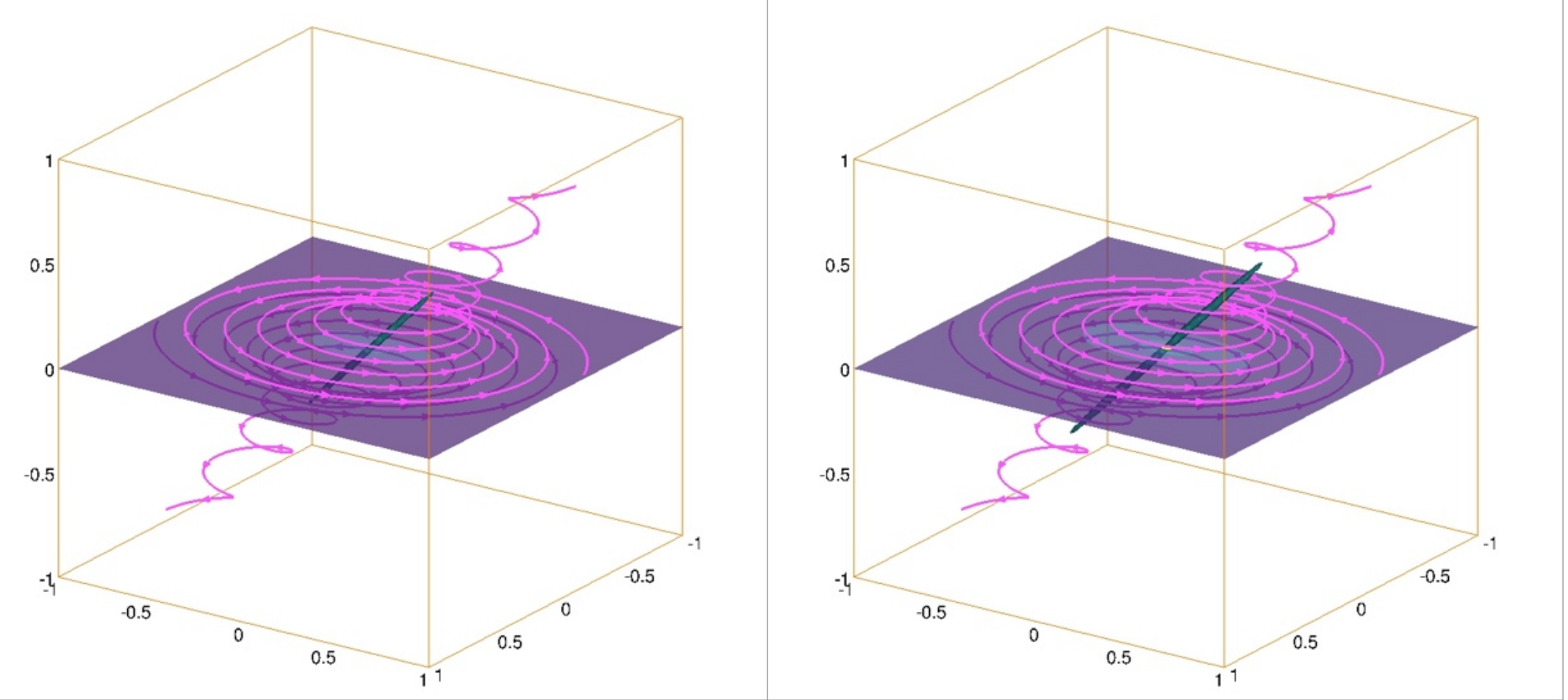

**Figure 5**. Iso-surfaces of $\mathcal{S}_d$ (left) and $\left|\mathrm{Int}_{\mathcal{L}}\, \widehat{\boldsymbol{A}}\right|$ (right) with value $0.5\pi$, where the normal $\boldsymbol{m}$ of $\mathcal{D}$ is chosed to be radial and the integration radius is 0.024. Field distributions show both $\mathcal{S}_d$ and $\left|\mathrm{Int}_{\mathcal{L}}\, \widehat{\boldsymbol{A}}\right|$ are less than $2\pi$.

In Fig. 6 and Table 2, we investigate the change of the topological indices $\mathcal{S}_d$ and $\left|\mathrm{Int}_{\mathcal{L}}\, \widehat{\boldsymbol{A}}\right|$ along different lines. Different formulae, namely the formulae (8) and $(38)_1$ directly from the definitions and, the derived formulae $(44)_3$ and (45) in the form of area integration and suitable only to $\mathcal{L}_{NC}$ (no value when $\boldsymbol{x}_0$ is the origin), are used. The results show the correctness of the derived formulae. All results show that $\mathcal{S}_d$ and $\left|\mathrm{Int}_{\mathcal{L}}\, \widehat{\boldsymbol{A}}\right|$ are less than $2\pi$, and $\mathcal{S}_d$ decrease faster than $\left|\mathrm{Int}_{\mathcal{L}}\, \widehat{\boldsymbol{A}}\right|$.

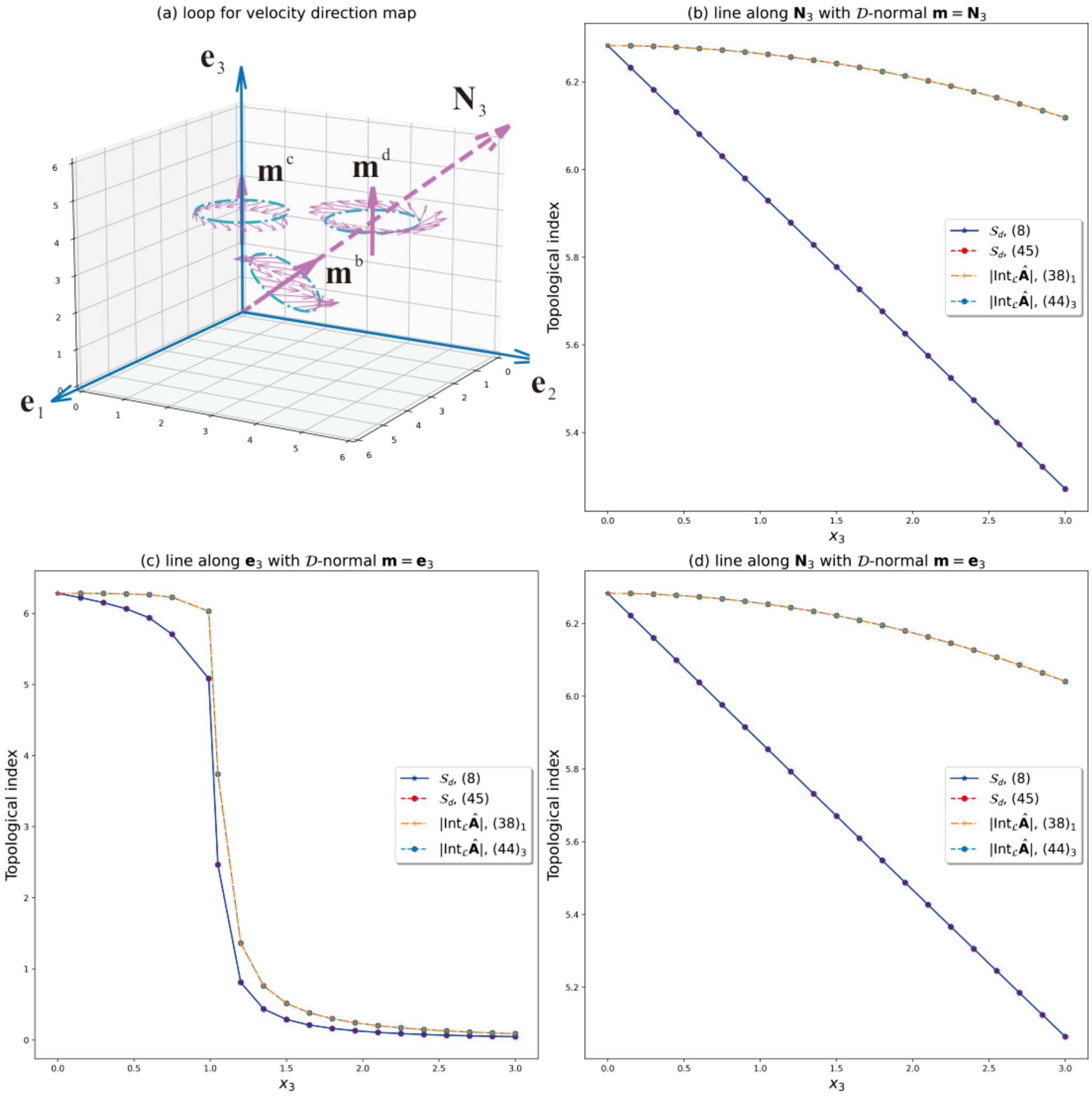

**Figure 6**. $\mathcal{S}_d$ and $\left|\mathrm{Int}_{\mathcal{L}}\,\widehat{\boldsymbol{A}}\right|$ along the $x_3$-axis and the extension line, where the integration radius is 1. (a) loop for velocity direction map, (b) line along $N_3$ with $\mathcal{D}$-normal $\boldsymbol{m} = \boldsymbol{N}_3$, (c) line along $\boldsymbol{e}_3$ with $\mathcal{D}$-normal $\boldsymbol{m} = \boldsymbol{e}_3$, (d) line along $\boldsymbol{N}_3$ with $\mathcal{D}$-normal $\boldsymbol{m} = \boldsymbol{e}_3$.

From Table 2 and Fig. 6(c), we see both $\mathcal{S}_d$ and $\left|\mathrm{Int}_{\mathcal{L}}\,\widehat{\boldsymbol{A}}\right|$ decrease suddenly after $x_3 = 1.05$, this is because the integration radius makes that $\mathcal{D}$ no longer intersects the extension line when the $x_3$ of $\boldsymbol{x}_0$ increases. The points on the extension line look like pseudo critical points. From Table 2, we can check an interesting result that $2\mathcal{S}_d > \left|\mathrm{Int}_{\mathcal{L}}\,\widehat{\boldsymbol{A}}\right|$ though $\mathcal{S}_d$ decreases rapider than $\left|\mathrm{Int}_{\mathcal{L}}\,\widehat{\boldsymbol{A}}\right|$.

**Table 2**. $\mathcal{S}_d$ and $\left|\mathrm{Int}_{\mathcal{L}}\,\widehat{\boldsymbol{A}}\right|$ along the $x_3$-axis with unit integration radius.

| $x_3$ | $\mathcal{S}_d$ | | $\left\|\mathrm{Int}_{\mathcal{L}}\,\widehat{\boldsymbol{A}}\right\|$ | |
|---|---|---|---|---|
| | (8) | (45) | $(38)_1$ | $(44)_3$ |
| 0.9 | 5.0826895 | 5.0826691 | 6.0300771 | 6.0300582 |
| 1.05 | 2.4652948 | 2.4651348 | 3.7391617 | 3.7388441 |
| 1.2 | 0.8097366 | 0.8097036 | 1.3611553 | 1.3610626 |
| 1.5 | 0.2880755 | 0.2880752 | 0.5149037 | 0.5148958 |
| 1.8 | 0.1620477 | 0.1620486 | 0.2991215 | 0.2991200 |
| 2.1 | 0.1070767 | 0.1070773 | 0.2016795 | 0.2016790 |
| 2.4 | 0.0770024 | 0.0770029 | 0.1470069 | 0.1470067 |
| 2.7 | 0.0584159 | 0.0584162 | 0.1125834 | 0.1125833 |
| 3.0 | 0.0460039 | 0.0460041 | 0.0892735 | 0.0892734 |

Introduce the module of the normal swirl field

$$\|\hat{\boldsymbol{A}}\| = \sqrt{\hat{A}_k^i \hat{A}_k^i}, \tag{54}$$

we illustrate its distribution in Fig. 7. It has dimension of reciprocal length, and also finds the extension line by its iso-surface with large value. However, $\|\hat{\boldsymbol{A}}\|$ has no definite topological explanation like $\left|\mathrm{Int}_{\mathcal{L}}\, \hat{\boldsymbol{A}}^i\right|$.

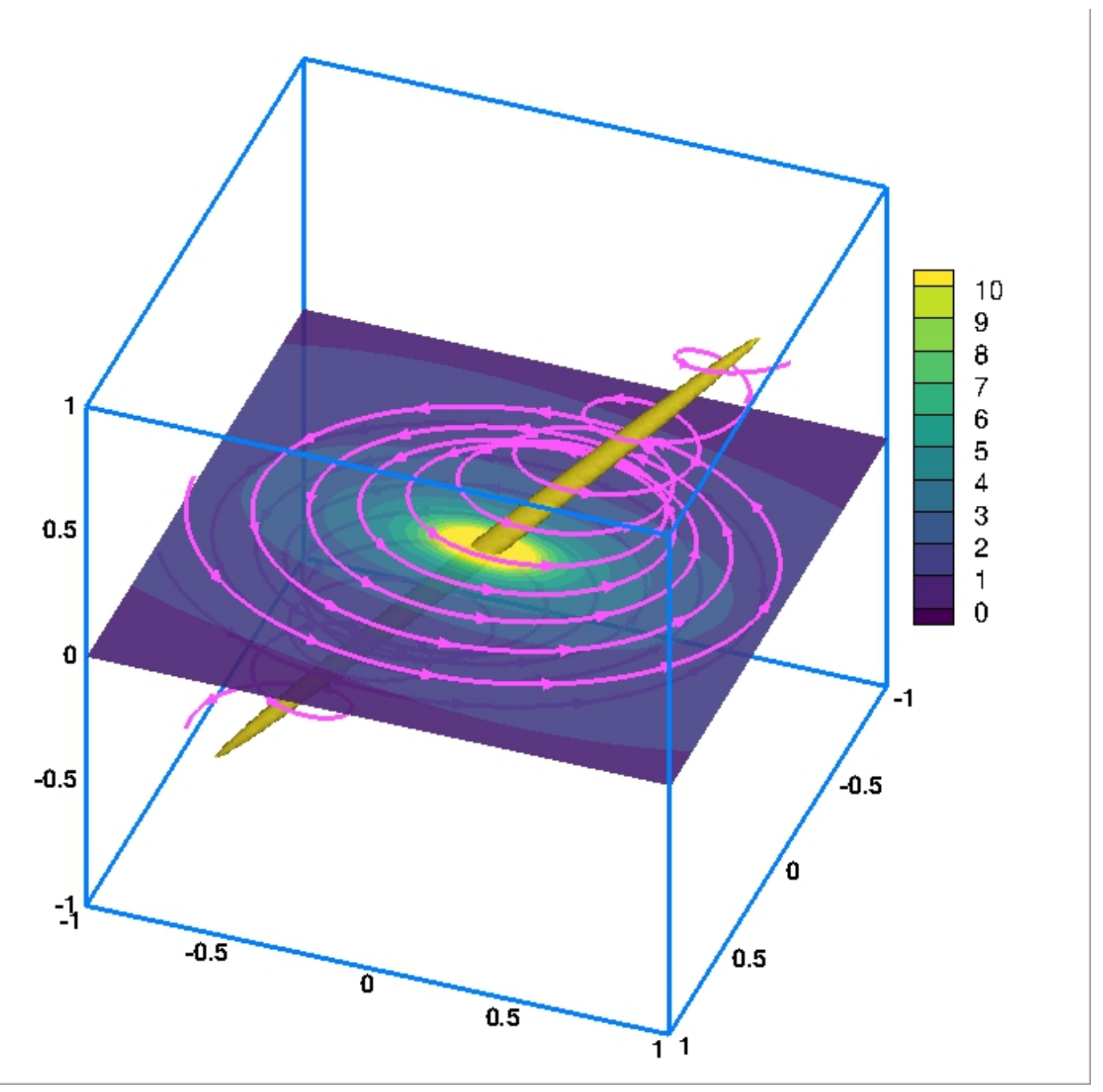

**Figure 7**. Module $\|\hat{\boldsymbol{A}}\|$ of the normal swirl field finds the extension line too, where the iso-surface value is 30.

### 5.2 Structures of non-spiral streamline pattern

All derivations above are independent of the sign of $\sigma$, but what we care about most up to now is the case of spiral streamlines under the condition $\sigma = -1$. In this subsection, we talk about the case $\sigma = 1$, where there are three real eigenvectors that are equally important. As pointed out in Zou *et al*. (2021), there are at least three base planes on which the streamlines are planar.

If there is a zero eigenvalue, we have a critical straight line along the corresponding left eigendirection, or equivalently the flow becomes two dimensional on the planes with the corresponding right eigendirection as their normal. In order to derive the critical lines of eigenvalues being zero, we do a rotation around $\boldsymbol{e}_3$

$$(\boldsymbol{e}_1^*, \boldsymbol{e}_2^*) = (\boldsymbol{e}_1, \boldsymbol{e}_2)\begin{pmatrix} \cos\gamma & -\sin\gamma \\ \sin\gamma & \cos\gamma \end{pmatrix}, (x_1^*, x_2^*) = (x_1, x_2)\begin{pmatrix} \cos\gamma & -\sin\gamma \\ \sin\gamma & \cos\gamma \end{pmatrix} \tag{55.1}$$

such that

$$\begin{pmatrix} \cos\gamma & \sin\gamma & 0 \\ -\sin\gamma & \cos\gamma & 0 \\ 0 & 0 & 1 \end{pmatrix}\begin{pmatrix} \vartheta - \frac{1}{2}\lambda_3 & \psi + \tau_3 & 0 \\ \psi & \vartheta - \frac{1}{2}\lambda_3 & 0 \\ \tau_1 & \tau_2 & \vartheta + \lambda_3 \end{pmatrix}\begin{pmatrix} \cos\gamma & -\sin\gamma & 0 \\ \sin\gamma & \cos\gamma & 0 \\ 0 & 0 & 1 \end{pmatrix} = \begin{pmatrix} \vartheta + \lambda_1 & \tau_3 & 0 \\ 0 & \vartheta + \lambda_2 & 0 \\ \tau_1^* & \tau_2^* & \vartheta + \lambda_3 \end{pmatrix} \tag{55.2}$$

with

$$\vartheta + \lambda_{1,2} = \vartheta - \tfrac{1}{2}\lambda_3 \pm \omega. \tag{55.3}$$

Direct calculation yields

$$\tan\gamma = \pm\sqrt{\frac{\psi}{\psi + \tau_3}}, \text{or } \cos 2\gamma = \frac{\tau_3}{2\psi + \tau_3}, \tag{55.4}$$

resulting in

$$\tau_1^* = \tau_1 \cos\gamma + \tau_2 \sin\gamma, \tau_2^* = \tau_2 \cos\gamma - \tau_1 \sin\gamma, \tag{55.5}$$

and the expression

$$\boldsymbol{v} = [(\vartheta+\lambda_1)x_1^* + \tau_1^* x_3]\boldsymbol{e}_1^* + [\tau_3 x_1^* + (\vartheta+\lambda_2)x_2^* + \tau_2^* x_3]\boldsymbol{e}_2^* + (\vartheta+\lambda_3)x_3\boldsymbol{e}_3. \tag{55.6}$$

The critical lines along the eigendirections and passing the origin are:

$$\begin{cases} \mathcal{S}_1 = \{\tau_3 x_1^* + (\vartheta+\lambda_2)x_2^* = 0, x_3 = 0\} & \text{, if } \vartheta+\lambda_1 = 0; \\ \mathcal{S}_2 = \{x_1^* = 0, x_3 = 0\} & \text{, if } \vartheta+\lambda_2 = 0; \\ \mathcal{S}_3 = \{(\vartheta+\lambda_1)x_1^* + \tau_1^* x_3 = 0, \tau_3 x_1^* + (\vartheta+\lambda_2)x_2^* + \tau_2^* x_3 = 0\} & \text{, if } \vartheta+\lambda_3 = 0. \end{cases} \tag{55.7}$$

It is obvious that for every situation above there is a unique critical line. For the cases with two zero eigenvalues, the corresponding critical lines are:

$$\begin{cases} \mathcal{S}_{12} = \{\tau_3 x_1^* = 0, x_3 = 0\} & \text{, if } \vartheta+\lambda_1 = \vartheta+\lambda_2 = 0; \\ \mathcal{S}_{23} = \{(\vartheta+\lambda_1)x_1^* + \tau_1^* x_3 = 0, \tau_3 x_1^* + \tau_2^* x_3 = 0\} & \text{, if } \vartheta+\lambda_2 = \vartheta+\lambda_3 = 0; \\ \mathcal{S}_{31} = \{\tau_1^* x_3 = 0, \tau_3 x_1^* + (\vartheta+\lambda_2)x_2^* + \tau_2^* x_3 = 0\} & \text{, if } \vartheta+\lambda_3 = \vartheta+\lambda_1 = 0. \end{cases} \tag{55.8}$$

The critical lines $\mathcal{S}_{12}$ and $\mathcal{S}_{31}$ could be extended to critical surfaces if $\tau_3 = 0$ and $\tau_1^* = 0$, respectively; another critical line $\mathcal{S}_{23}$ is also possible to be extended to a critical surface on certain conditions. For the case with three zero eigenvalues, the critical line is

$$\mathcal{S}_{123} = \{\tau_1^* x_3 = 0, \tau_3 x_1^* + \tau_2^* x_3 = 0\}, \text{ if } \vartheta+\lambda_1 = \vartheta+\lambda_2 = \vartheta+\lambda_3 = 0, \tag{55.9}$$

which could be extended a surface or even the whole space if $\tau_3 = \tau_2^* = \tau_1^* = 0$. In summary, we have no more than four simple situations according to the zero-value property of three real eigenvalues of a linear velocity: one critical point, one critical line, one zero plane and the whole zero. The last situation is trivial with no need to discuss.

Back to the topological indices $\mathcal{S}_d$ and $\left|\mathrm{Int}_{\mathcal{L}}\,\widehat{\boldsymbol{A}}\right|$, we first consider cases of one zero plane and one critical line. Since $\det(\boldsymbol{D}) = 0$, we have $\mathcal{S}_d = \left|\mathrm{Int}_{\mathcal{L}}\,\widehat{\boldsymbol{A}}\right| = 0$ from (44) and (45) on all $\mathcal{L}_{NC}$. For the loop $\mathcal{L}_C$:

- One zero plane: using (50) for $\mathcal{L}_C$ intersecting the plane ($\boldsymbol{N}_3 \cdot \boldsymbol{m} = 0$), and the null velocity on $\mathcal{L}_C$ when it is on the critical plane, we always have $\mathcal{S}_d = \left|\mathrm{Int}_{\mathcal{L}}\,\widehat{\boldsymbol{A}}\right| = 0$; so, zero plane is a trivial topological structure!
- One critical line: besides the critical line on the plane of $\mathcal{L}$ such that $\mathcal{S}_d = \left|\mathrm{Int}_{\mathcal{L}}\,\widehat{\boldsymbol{A}}\right| = 0$; from (48) we have $\mathrm{Int}_{\mathcal{L}_C}\,\widehat{\boldsymbol{A}} = -2\pi\widehat{\boldsymbol{k}}$ where the vector $\boldsymbol{k}$ can be expressed by the parameters in the new coordinate system $(x_1^*, x_2^*, x_3)$ as
$$\boldsymbol{k} = (\boldsymbol{e}_1^* \quad \boldsymbol{e}_2^* \quad \boldsymbol{e}_3)\begin{pmatrix} (\vartheta+\lambda_3)(\vartheta+\lambda_2) & -\tau_3(\vartheta+\lambda_3) & 0 \\ 0 & (\vartheta+\lambda_3)(\vartheta+\lambda_1) & 0 \\ -\tau_1^*(\vartheta+\lambda_2) & \tau_1^*\tau_3 - \tau_2^*(\vartheta+\lambda_1) & (\vartheta+\lambda_1)(\vartheta+\lambda_2) \end{pmatrix}\begin{pmatrix} m_1^* \\ m_2^* \\ m_3 \end{pmatrix}. \tag{56}$$
The new formula is capable of considering different cases of eigenvalues becoming zero. And numerically, we test $\mathcal{S}_d = 2\pi$ for all $\mathcal{L}_C$.
- Two repeated roots $\lambda_1 = \lambda_2 = \lambda$, besides $\boldsymbol{e}_3$: (1) there is an intrinsic direction $(\lambda - \lambda_3)\boldsymbol{e}_1^* + \tau_3\boldsymbol{e}_3$ if $\tau_3 \neq 0$; (2) there are infinite intrinsic directions in the plane normal to $-\tau_1^*\boldsymbol{e}_1^* - \tau_2^*\boldsymbol{e}_2^* + (\lambda - \lambda_3)\boldsymbol{e}_3$ if $\tau_3 = 0$.
- Three repeated roots $\lambda_1 = \lambda_2 = \lambda_3 = \lambda$, besides $\boldsymbol{e}_3$ (i) $\tau_3 \neq 0$, there is no intrinsic direction if $\tau_1^* \neq 0$, or there are infinite intrinsic directions in the plane normal to $\boldsymbol{e}_2^*$ if $\tau_1^* = 0$; (ii) $\tau_3 = 0$, there are infinite intrinsic directions in the plane normal to $\tau_1^*\boldsymbol{e}_1^* + \tau_2^*\boldsymbol{e}_2^*$, or all directions are intrinsic if $\tau_3 = \tau_1^* = \tau_2^* = 0$.

About the situation without zero eigenvalue, Fig. 8 shows there is a left real eigendirection and the unit sphere mapping of linear velocity with gradient

$$\boldsymbol{D} = \begin{bmatrix} -0.001 & 0.009 & 0 \\ 0.004 & -0.001 & 0 \\ 0.001 & -0.003 & 0.002 \end{bmatrix},$$

and with the origin as its critical point. The topological index $\left|\mathrm{Int}_{\mathcal{L}}\,\widehat{\boldsymbol{A}}\right|$ shown in Fig. 9 concentrates along the eigendirection of eigenvalue with the minimum modulus, where the iso-surface is no longer like a needle with two sharp ends but like a spindle. Another topological index $\mathcal{S}_d$ is similar but the spindle-like iso-surface is smaller. These results indicate that there is still an extension direction though it is not as obvious as spiral streamline.

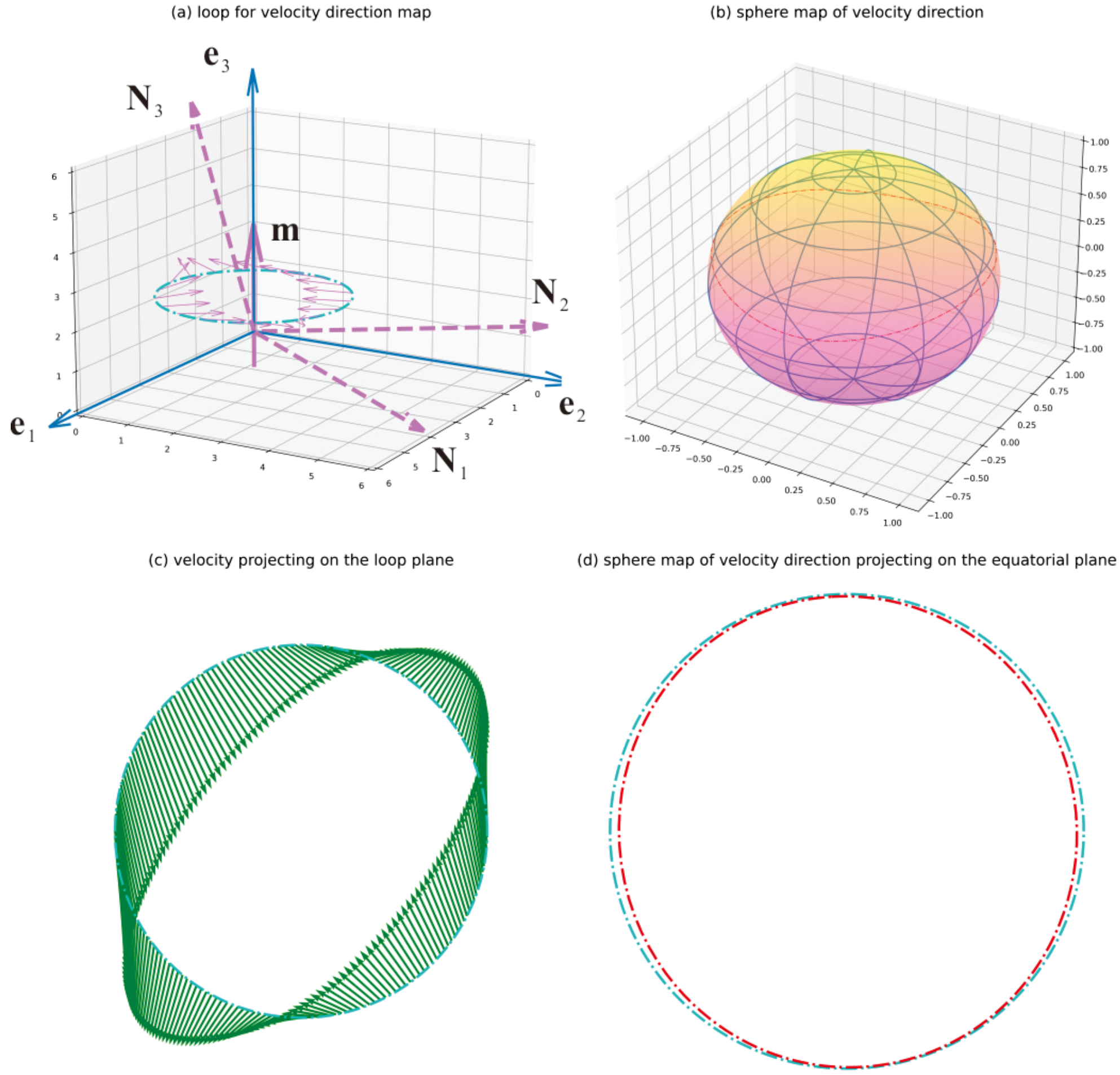

**Figure 8**. Loop and mapping onto unit sphere (three real eigenvalues case): (a) loop for velocity direction map, (b) sphere map of velocity direction, (c) velocity projecting on the loop plane, (d) sphere map of velocity direction projecting on the equatorial plane.

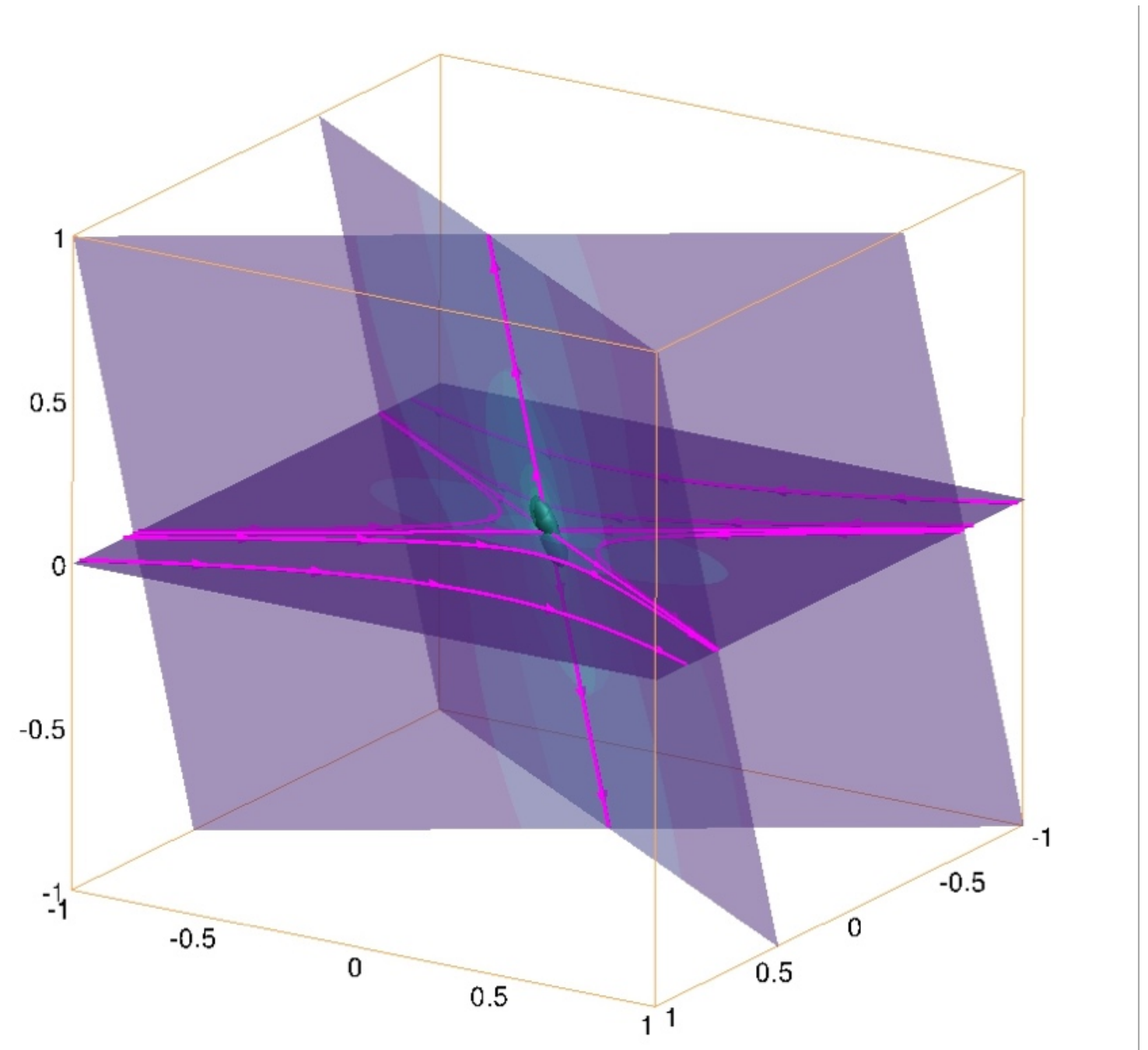

**Figure 9**. Iso-surfaces of $\left|\mathrm{Int}_{\mathcal{L}}\, \widehat{\boldsymbol{A}}\right|$ with value $0.5\pi$, where the $\mathcal{D}$-normal $\boldsymbol{m}$ is radial and the integration radius is 0.024. The concentration of $\left|\mathrm{Int}_{\mathcal{L}}\, \widehat{\boldsymbol{A}}\right|$ is along the eigendirection whose eigenvalue has minimal module.

### 5.3 About the definition of orientation frame

The velocity direction is no doubt the most important factor of fluid transporting in steady flows. The determination of orientation frame depends on the dynamical process of fluid. Here we provide two choices to be tested in practice:

(1) One is of kinematic (Fig. 10), we define the plane of shear by velocity direction $\mathbf{n}_1$ and speed gradient $\nabla V$ so that

$$\mathbf{n}_1 \times \nabla V = \Pi \mathbf{n}_3, \qquad \Pi = |\mathbf{n}_1 \times \nabla V|. \tag{57.1}$$

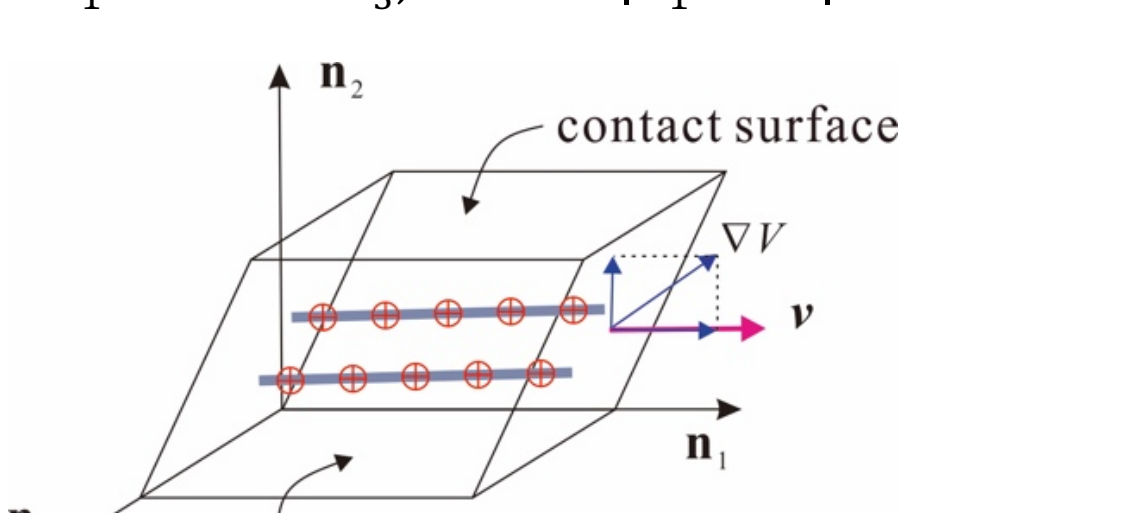

**Figure 10.** Shear flow and contact surface

The operable formulae for $\Pi$ and the swirl field $\boldsymbol{A}$ can be derived to be

$$\Pi = \frac{|V^2 \nabla V^2 - (\boldsymbol{v} \cdot \nabla V^2)\boldsymbol{v}|}{2V^3}, \tag{57.2}$$

$$A_k^l \mathbf{e}_l dx_k = -\frac{\boldsymbol{v} \times d\boldsymbol{v}}{V^2} + \left(\frac{\boldsymbol{v} \times \nabla V^2}{2V^2 \Pi} \cdot \frac{d\boldsymbol{v}}{V}\right) \frac{\boldsymbol{v} \cdot \nabla V^2}{2V^2 \Pi} \frac{\boldsymbol{v}}{V} - \left(\frac{\boldsymbol{v} \times \nabla V^2}{2V^2 \Pi} \cdot \frac{d\nabla V^2}{2V\Pi}\right) \frac{\boldsymbol{v}}{V}. \tag{57.3}$$

This seems to be a good choice when there is evidence that shearing flow may induce fluid layering, but it is awful to know that some distribution of speed can form a streamwise vortex.

(2) Another is of kinetic (Fig. 3), we get the plane of force (the explanation will be a part of another paper on the new dynamics of fluid flow) by velocity direction $\mathbf{n}_1$ and streamline bending $\mathbf{n}_1 \cdot \nabla \mathbf{n}_1$, so that

$$\mathbf{n}_1 \cdot \nabla \mathbf{n}_1 = \kappa \mathbf{n}_2, \qquad \kappa = |\mathbf{n}_1 \cdot \nabla \mathbf{n}_1|. \tag{58.1}$$

The operable formulae for the curvature $\kappa$ and the swirl field $\boldsymbol{A}$ can be derived to be

$$\kappa = |\mathbf{n}_1 \cdot \nabla \mathbf{n}_1| = \frac{\left|V^2 \boldsymbol{v} \cdot \nabla \boldsymbol{v} - \frac{1}{2}(\boldsymbol{v} \cdot \nabla V^2)\boldsymbol{v}\right|}{V^4}, \tag{58.2}$$

$$A_k^l \mathbf{e}_l dx_k = -\frac{\boldsymbol{v} \times d\boldsymbol{v}}{V^2} + \left[\frac{\boldsymbol{v} \cdot \nabla V^2}{2\kappa V^3} \frac{\boldsymbol{v} \times (\boldsymbol{v} \cdot \nabla \boldsymbol{v})}{\kappa V^3} \cdot \frac{d\boldsymbol{v}}{V} - \frac{\boldsymbol{v} \times (\boldsymbol{v} \cdot \nabla \boldsymbol{v})}{\kappa V^3} \cdot \frac{d(\boldsymbol{v} \cdot \nabla \boldsymbol{v})}{\kappa V^2}\right] \frac{\boldsymbol{v}}{V}. \tag{58.3}$$

This choice is more geometrical, and may indicate the pressure effect when connecting with the advection term in the momentum equation.

We prefer the latter choice due to its pure geometric attribute in this paper. This makes the definition of swirl field more descriptive. Covariant differential (17) of velocity indicates the way the swirl field couples with the velocity field. In steady flows, this minimal coupling means the stable orientation difference of fluid is not responsible for the non-uniformity of flow, namely

$$D\boldsymbol{v} = d\boldsymbol{v} + \boldsymbol{A} \times \boldsymbol{v} = (dV)\boldsymbol{n}_1. \tag{59}$$

But when the flow becomes unstable or turbulent, this coupling becomes a key means for the swirl field as an independent field to affect the flow heterogeneity, that is, to affect the viscous interaction, and is no longer restrained by the critical point (line) of velocity field. The detailed mechanisms of the swirl field, together with the eddy field, will be described in our next article.

## 6. Conclusions

In this paper, combined with the slip model of real fluid flows, we study the topological description of streamline pattern around the critical point, with the linear velocity fields as an example. Two concepts, the swirling degree of velocity field and swirl field, are introduced to characterize the topology of flow, their operable formulae are derived. In steady flows, the swirl field is defined by the difference of frames which are connected with the velocity direction and maybe the gradient of speed. For planar flows, the swirling degree, with values of integer multiples of $2\pi$, is a perfect classification criterion to streamline pattern. But It cannot be simply generalized to three-dimensional flows. Through derivation, demonstration and numerical test based on linear velocity field, the following conclusions are obtained:

(1) $\mathcal{S}_d$ and $|\mathrm{Int}_{\mathcal{L}}\,\widehat{\boldsymbol{A}}|$ are almost equivalent as a classification criterion of structures, but $\mathrm{Int}_{\mathcal{L}}\,\widehat{\boldsymbol{A}}$ and $\boldsymbol{A}$ have richer intensions;

(2) The kinds of singularity structures of $\mathcal{S}_d$ and $|\mathrm{Int}_{\mathcal{L}}\,\widehat{\boldsymbol{A}}|$ include only single point and single line, no plane and volume;

(3) The critical line can be identified cleanly by $\mathcal{S}_d$ and $|\mathrm{Int}_{\mathcal{L}}\,\widehat{\boldsymbol{A}}|$, while the critical point can be identified together with an extension direction, but the morphology and strength of iso-surface are quite different for spiral streamline and non-spiral streamline;

(4) The dual directivity built up from right eigen-representation is also the property of $\mathcal{S}_d$ and $|\mathrm{Int}_{\mathcal{L}}\,\widehat{\boldsymbol{A}}|$.

In the generalization from 2D to 3D, the integral loop involves not only shape, size, but also directionality, the swirling degree is no longer integer multiples of $2\pi$. This brings a lot of confusion. According to the study in this paper, we conclude that the nonzero of loop integration of the swirl field around a regular point comes from the non-commutativity of 3D rotation group, which can be made up by an additional nonlinear term like (51). And the whole expression (53.5) is independent of the choice of shear plane. If the addition integration from the nonlinear term (51) is considered, we reobtain the property of swirl field to distinguish the singular point of orientation from the regular ones. The value of generalized integration is a vector dependent on the normal of loop surface, but its amplitude is an integer multiple of $2\pi$. This means that the sign of $\det(\boldsymbol{D})$ is not important as in 2D flows.

How about the nonlinear velocity fields? We know for 2D nonlinear flows the classification of streamline pattern around an isolated point has to be more complex (Jiang and Llibre, 2005; Artés et al., 2015, 2021). Besides the swirling degree, more meticulous concepts such as separatrices, sectors and their types are introduced to make a distinction. The combination of different types of critical points to form a complex flow structure becomes quite common. In 3D flows, a quadratic field can exhibit chaotic-like behavior, called the strange attractor (Lorenz, 1963; Ruelle and Takens, 1971). The pattern classification research of 3D nonlinear flows near a critical point still has a long way to go.

When the flows become unsteady or even turbulent, the streamline description is lost since there may be no zero point of velocity. In the viewpoint of slip model, it is the singularity of velocity direction instead of the singularity of velocity itself that creates the complexity of streamline pattern, and the evolution of the vortex field independent of the macroscopic velocity achieves untrained turbulent flows. The singularity of velocity direction, more than the various of vortices, is the key to unlocking the mystery of turbulence.

In most vortex identification methods, linear velocity field achieves a uniform structure, and few researches realize the local character of structure as a point or a line, and insist on the presence of a critical point to be necessary for a structure. In addition, the structure identified by the swirling degree is geometrical instead of a physical, the strength of a structure makes no sense.